\documentclass[lettersize,journal]{IEEEtran} %
\usepackage{amsmath,amsfonts}
\usepackage{algorithmic}
\usepackage{algorithm}
\usepackage{array}
\usepackage[caption=false,font=normalsize,labelfont=sf,textfont=sf]{subfig}
\usepackage{textcomp}
\usepackage{stfloats}
\usepackage{url}
\usepackage{verbatim}
\usepackage{graphicx}
\usepackage{cite}
\begin{document}

\title{Semantic Bundling: Interactive Node and Edge Bundling to Simplify Knowledge Graphs using Large Language Models}

\author{Adam Coscia, Zeyu Hua, Eric Krokos, Timothy Lin, Alex Endert
\thanks{A. Coscia, Z. Hua, and A. Endert are with School of Interactive Computing, Georgia Institute of Technology. E. Krokos and T. Lin are with DOD. }%
}%

\maketitle

\begin{abstract}
  We present Semantic Bundling, a visual analytics technique for making sense of text documents represented as knowledge graphs (KGs). 
  Representing a document corpus as a KG makes relationships between entities explicit, making KGs useful both to analyze directly and in computational workflows including ML pipelines and generative AI backends.
  However, as KGs grow they become difficult to interpret and visualize for specific tasks (e.g., the ``hairball problem''), with the meaning of each relationship often buried in dense source text.
  Semantic Bundling uses large language models (LLMs) to support user-driven bundling of nodes and edges in a KG into higher-level graph structures: \textit{super nodes}, which collapse and summarize a region of the graph, and \textit{super edges}, which summarize the connection between two entities.
  Results are linked to underlying triples and source documents, grounding summaries in evidence.
  We implement Semantic Bundling in AgentK, an open-source system that builds a KG from text documents and maps graph interactions to bundling operations. 
  Through use cases on movie reviews and an intelligence analysis scenario, we show how Semantic Bundling reveals new insights in document collections, and synthesize our findings into a discussion of emerging challenges and opportunities in knowledge graph sensemaking.
\end{abstract}


\section{Introduction}
\label{sec:introduction}

\IEEEPARstart{M}{aking} sense of a large collection of text documents often means understanding the relationships among the entities within it; i.e., who is connected to whom, and how.
Reading every document to reconstruct these relationships is time-consuming and error-prone, as the relationships often hide in dense passages and span across many documents.
One method to address this limitation is to represent the documents as a knowledge graph (KG), making the relationships latent in the text explicit as a structure of entities and the connections among them~\cite{abdurahman2021lex2kg,stylianou2022doc2kg,szekely2015building}.
For example, analysts, engineers, and scientists increasingly turn collections of reports, articles, and papers into knowledge graphs of people, organizations, and events to reason over how they relate, using tools like Obsidian \cite{Obsidian}, Atlas \cite{Atlas}, and Neo4j's KG Builder \cite{Neo4jKGBuilder}.

Yet supporting direct manipulation and exploration of KGs during visual data analysis remains challenging.
As KGs grow, they quickly devolve into dense "hairballs" that obscure key relationships. This compounds the difficulty of core tasks such as filtering parts of the graph, tracing relationships across multiple hops, and linking relationships back to the source text to validate findings.
Document-centric visual analytics keeps analysis in the space of text, leaving relationships implicit through the use of topic modeling~\cite{havre2002themeriver}, clustering~\cite{wise1995visualizing}, or a combination~\cite{Stasko:2007:Jigsaw,endert2012semantic, Coscia:2026:VisPile} to handle documents at scale.
Graph and knowledge graph visualizations instead make relationships explicit, focusing mainly on presenting the graph structure in an interpretable and interactive way~\cite{chau2011apolo,di2021stratisfimal}.
However, analyzing the structure of relationships extracted from text documents, while simultaneously making the KG topology human-readable and useful, remains underexplored.
Our work investigates using large language models (LLMs) to synthesize and generate intermediate representations that can bridge this gap during visual data analysis.

We present \textbf{Semantic Bundling}, a visual analytics technique for making sense of text documents represented as KGs.
Semantic Bundling comprises two operations: \textit{super nodes}, which collapse and summarize a region of the graph, and \textit{super edges}, which summarize the connection between two entities.
Users can generate semantic bundles to relabel their KG through direct graph manipulation, bridging the visual graph exploration process with operations that make the KG easier to parse at glance.
Semantic bundles also map the generated labels and summaries to their underlying nodes and edges, making bundles traceable back to source documents.

We realize Semantic Bundling in \textbf{AgentK}, an open-source\footnote{URL: [blinded for review]} proof-of-concept system that builds a knowledge graph from documents, maps graph interactions to Semantic Bundling operations, and visualizes the results for inspection.
To validate our approach, we present use cases on the IMDb movie reviews dataset~\cite{zm1y-b270-20} and an intelligence analysis scenario with the Kronos dataset~\cite{whiting2014vast}, showing how users uncover and trace insights while keeping the source documents a click away.
Finally, we discuss open gaps in supporting KG sensemaking more broadly, as well as the trade-offs we encountered in letting an LLM define the graph's ontology.

In summary, our paper contributes:
\begin{enumerate}
    \item \textbf{Semantic Bundling}, a visual analytics technique comprising two LLM-driven operations over a knowledge graph: \textbf{super nodes} and \textbf{super edges}, which abstract, summarize, and connect relationships while staying traceable to their source evidence.
    \item \textbf{AgentK}, an open-source proof-of-concept visual analytics system implementing Semantic Bundling for knowledge graph analysis.
    \item \textbf{Usage scenarios} demonstrating how Semantic Bundling reveals insights in two types of document corpora.
\end{enumerate}

\section{Related Work}
\label{sec:related_work}

Semantic Bundling draws on visual analytics for document sensemaking, graph and knowledge graph visualization, and the intersection of LLMs with knowledge graphs.

\subsection{Visual Analytics for Document Sensemaking}

Visual analytics has long helped people make sense of large text collections, surfacing the entities and relationships buried across many documents to support the sensemaking loop in investigative analysis \cite{Pirolli:2005:SensemakingProcess}.

Early work established spatial encoding as a primary metaphor for navigating large corpora.
For example, Galaxy~\cite{wise1995visualizing} projects documents into a 2D plane according to lexical similarity, allowing users to perceive topical clusters at a glance without reading individual reports.
Entity-centric systems shift the unit of analysis from documents to the named people, places, organizations, and events they contain.
Jigsaw~\cite{Stasko:2007:Jigsaw} embodies this approach with a coordinated multi-view design that makes cross-document co-appearances of entities explicit, letting users trace threads of evidence through synchronized list, graph, scatterplot, and text views.

More recent work \cite{endert2012semantic, Coscia:2026:VisPile} turns to machine learning to expose deeper semantic structure within document collections.
Ji et al.~\cite{ji2019visual} address the opacity of neural document embeddings by rendering them as a configurable document map, through which users can explore the embedding space, identify semantically salient neural dimensions, and perform guided feature selection to improve Information Retrieval (IR) performance.
ThoughtFlow~\cite{guo2018topic} applies topic modeling to literature corpora to support the ideation stage of research, enabling exploratory browsing across thematic clusters while embedded in-situ visualizations help users assess document relevance without losing their place.
A common theme among these tools is grounding analysis in the space of documents and text; however, the relationships among entities remain implicit, leaving the user to reconstruct them as first-class structures.
We instead move analysis onto the knowledge graph itself, making relationships the primary object of interaction.

\subsection{Graph and Knowledge Graph Visualization}

Most systems visualize KGs as node-link diagrams.
Interactive exploration addresses part of the scale challenge by letting users navigate incrementally rather than confronting the full graph at once.
Apolo~\cite{chau2011apolo} exemplifies this with a mixed-initiative approach that combines Belief Propagation and rich user interaction to infer relevant nodes from user-specified exemplars, guiding bottom-up sensemaking of large networks.
For structured graphs, layout quality is itself a research problem, and STRATISFIMAL layout~\cite{di2021stratisfimal} formulates node placement in layered node-link diagrams as a modular optimization, balancing crossing reduction, compactness and alignment.
Researchers have also applied KGs as infrastructure for other visualization tasks, as KG4Vis~\cite{li2021kg4vis} encodes relationships between data attributes and chart types in a knowledge graph and learns embeddings over it to recommend effective visualizations for new datasets.
Immersive environments offer yet another strategy for managing spatial complexity, with Kellman et al.~\cite{kellmann2024visualization} enabling users to navigate RDF-linked data in virtual reality rather than compressing it onto a flat display.

To cope with scale, prior work reduces visual clutter through edge bundling~\cite{Sun:2016:BiSet}, motif simplification~\cite{dunne2013motif}, and clustering or aggregation~\cite{elmqvist2009hierarchical}.
These techniques simplify graph \emph{structure} but do not convey the \emph{meaning} of what they collapse.
Semantic Bundling builds on this line of work, but uses an LLM to label and summarize collapsed regions and paths, so that simplification carries semantics rather than discarding it.

\subsection{Large Language Models and Knowledge Graphs}
A growing body of work pairs LLMs with knowledge graphs.
Retrieval-augmented approaches use a KG as a backend to ground LLM question answering.
Natural language (NL) interfaces then help users query KGs without writing formal queries.
Pan et al.~\cite{pan2024unifying} survey this landscape as a roadmap, organizing approaches by whether they use KGs to enhance LLM reasoning, use LLMs to augment construction and completion, or tightly couple both in a synergized system.
Think-on-Graph~\cite{sun2024think} exemplifies the retrieval-augmented direction, treating the LLM as an agent that iteratively searches the KG for supporting evidence, reducing hallucination by grounding each reasoning step in graph-retrieved paths.

In visual analytics more broadly, LLMs increasingly automate and assist analysis \cite{Wang:2024:AVA}, while human-in-the-loop approaches stress that people must remain able to verify model output.
Hutchinson et al.~\cite{hutchinson2024llm} identify reliability and user verification of model output as central open problems in LLM-assisted visual analytics.
LEVA~\cite{zhao2024leva} addresses this by weaving LLMs into the onboarding, exploration and summarization stages of a visual analytics workflow, while exposing an interaction history so users can retrace and audit LLM-generated insights.
These systems largely treat the knowledge graph as a retrieval or question-answering backend behind a text interface.
We instead place the knowledge graph in front as the analytic surface, exposing LLM operations through direct graph interaction and keeping their results traceable to the underlying evidence.

\section{Semantic Bundling Technique}
\label{sec:technique}

We present Semantic Bundling, a visual analytics technique for making sense of text documents represented as knowledge graphs (KGs).
Semantic Bundling transforms a KG into aggregated structures and labels that can improve sensemaking and analysis of the KG as well as the information it represents.
Our technique uses large language models (LLMs) to label and summarize groups of relationships (edges) and entities (nodes) while preserving links back to the underlying triples and source documents.
The technique comprises two categories of operations: \textit{super nodes}, which collapse and summarize a region of the graph, and \textit{super edges}, which summarize the connection between two entities.

In this section, we first scope the technique and define the KG terminology we use (Sect.~\ref{sec:sb/scope}).
We then describe the design challenges that motivate Semantic Bundling, drawn from observations of the knowledge graph, visualization, and sensemaking literature (Sect.~\ref{sec:sb/challenges}).
Finally, we detail each operation in turn (Sect.~\ref{sec:sb/super_nodes} and Sect.~\ref{sec:sb/super_edges}).

\subsection{Scope, Tasks, and Terminology}
\label{sec:sb/scope}

A \textit{knowledge graph} (KG) represents a document collection as a set of \textit{triples}, each written $\langle$\textit{subject, predicate, object}$\rangle$ and asserting a single relationship between two entities; e.g., $\langle$\textit{GAStech, operates in, Kronos}$\rangle$.
We refer to the entities as \textit{nodes} and the relationships between them as \textit{edges}, and many edges may join a single pair of nodes.
We scope our technique to work within \textit{closed document sets}: fixed, pre-curated collections that are too large to read in full but bounded in size, on the order of a thousand documents or less, such as the reports gathered for a single investigation.
We assume these documents have been, or can be, converted into a knowledge graph; large collections of documents are increasingly available in this form~\cite{abdurahman2021lex2kg,stylianou2022doc2kg,szekely2015building}.
For example, a user may want to understand what role a particular entity plays, trace how two entities are connected, or get a thematic overview of a densely-connected subgraph.

\subsection{Design Challenges}
\label{sec:sb/challenges}

Representing a document collection as a knowledge graph makes its relationships explicit, but the resulting graph is difficult to read and manipulate directly.
We observe four recurring challenges in visualizing and making sense of KGs that motivate the design of Semantic Bundling.

\medskip
\noindent\textbf{C1 -- Graph structure is hard to visually parse.}
Node-link diagrams are the dominant representation for knowledge graphs, but as the number of entities and relationships grows, they often degrade into dense ``hairballs'' that visually bury the relationships, groups, and patterns a user is looking for.
For example, a single highly-connected entity such as a major organization can accumulate hundreds of incident edges, collapsing its neighborhood into an unreadable knot.
A large body of work reduces this clutter through motif simplification~\cite{dunne2013motif} and hierarchical aggregation or sampling~\cite{elmqvist2009hierarchical}.
Yet these techniques operate mainly on the graph structure, which can cause the description of what was collapsed to become vague or hard to trace.
In other words, simplifying the substructures of a complex graph can create a representation that is easier to visually parse, but it does not describe what the collapsed region is about or what comprises it.
A solution should instead disentangle hairballs by labeling semantically similar regions that map back to the original graph structure.

\medskip
\noindent\textbf{C2 -- Relationships between two nodes can be complex to interpret.}
KGs often contain many relationships between the same two adjacent entities.
For example, these relationships may be fine-grained edges that together describe a higher-level phenomenon (e.g., the many contacts between two companies that indicate they work together), semantically redundant statements of the same fact (e.g., one person being both the ``mother of'' and the ``parent of'' another), or even contradictions (e.g., outdated edges, or conflicts introduced by merging datasets).
Prior work bundles such edges by structure to reduce clutter~\cite{Sun:2016:BiSet}, but does not describe what they mean together.
Giving users a way to semantically aggregate and label these relationships would let them quickly grasp an otherwise tangled connection between two entities.

\medskip
\noindent\textbf{C3 -- Connections between distant entities are hard to trace and interpret.  }
A central analytic task over a knowledge graph is to determine how two entities are connected; e.g., ``how are these two people related?''
Answering it requires following one or more multi-hop paths between the entities.
However, the number and length of these paths quickly becomes untenable for manual tracing.
For example, recent work has the LLM itself traverse the graph hop by hop to reason over such paths~\cite{sun2024think}, but this surfaces the chains for the model, not for a person reading the graph.
Existing path visualizations likewise reveal which paths exist, yet leave the user to assemble an interpretable account of what the connection means from the raw chain of relationships.
A solution should label the logical chain of connections at a higher level, letting users more easily understand the connection between distant nodes.

\medskip
\noindent\textbf{C4 -- Automated summarization can become ungrounded.  }
LLMs can summarize a subgraph into fluent natural language while grounding reasoning in graph-retrieved evidence -- a known way to reduce hallucination~\cite{sun2024think}.
But fluency can be misleading.
LLMs will still produce ungrounded synthesis, asserting connections the underlying data does not support.
Verifying their output remains an open problem for LLM-assisted visual analytics~\cite{hutchinson2024llm}.
For example, an LLM asked to summarize a subgraph may infer a plausible motive linking two people that no edge or document actually supports.
Sensemaking in high-stakes settings such as intelligence analysis instead demands traceability, where every generated claim must be attributable to the evidence it came from.
A technique that summarizes a knowledge graph with an LLM must therefore keep its output linked to the associated triples and documents, and use these to guide the generation process.

\medskip
Semantic Bundling addresses these challenges through two operations.
Super nodes target the node structure complexity (C1) by collapsing a region of the graph into a single summarized, structurally recognizable glyph motif.
Super edges target the connection tracing challenges (C2, C3) by summarizing how two entities are linked.
We design both operations to keep their generated summaries grounded in the source data (C4), described at the end of this section.

\begin{figure*}[!t]
  \includegraphics[width=\textwidth]{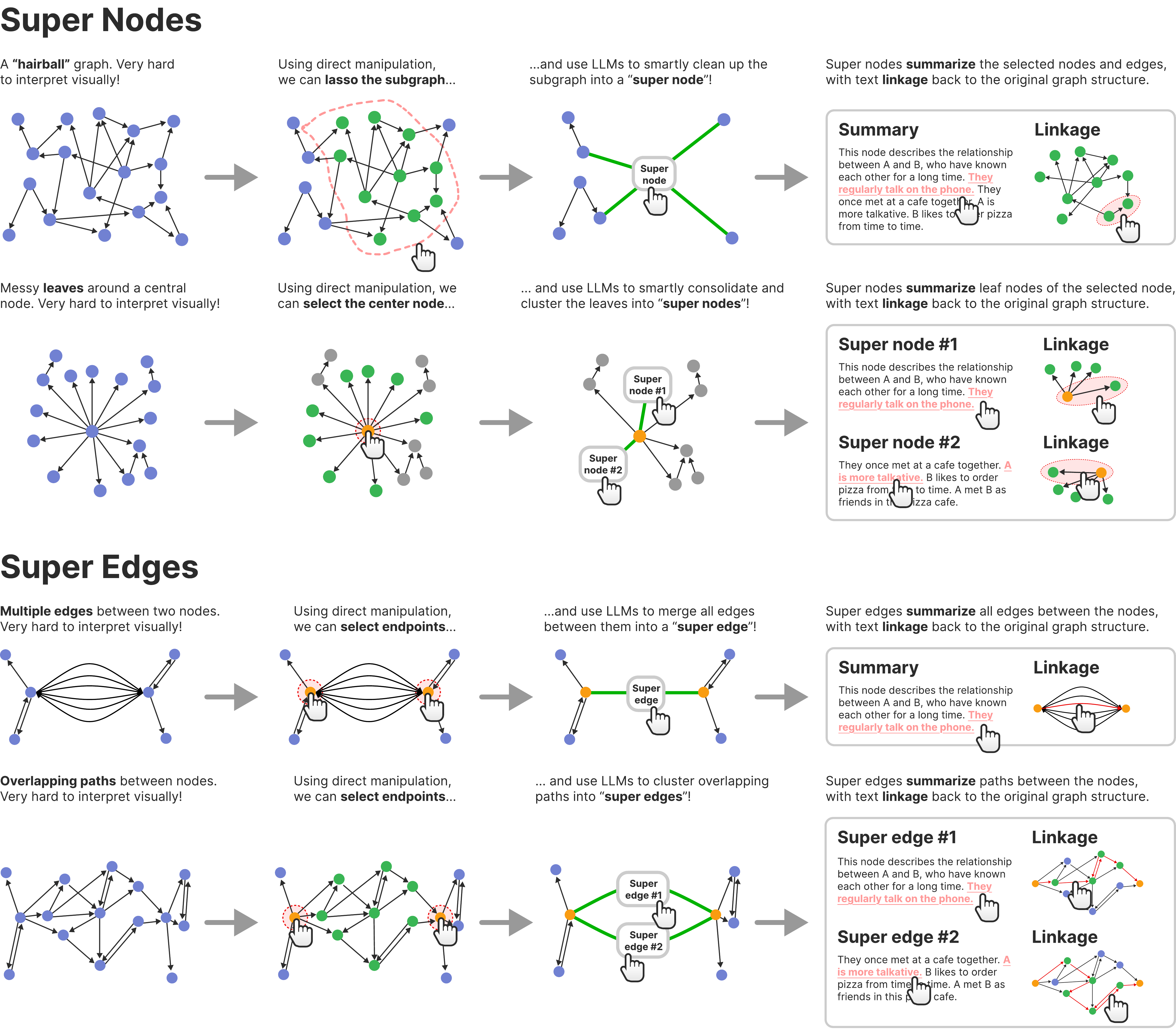}
  \caption{Semantic Bundling maps a user's graph selection to an operation that collapses the selected structure into a labeled, summarized glyph. Selecting a single node or lassoing a subgraph produces a \textit{super node} (top), which summarizes a region of the graph; selecting two adjacent or two distant entities produces a \textit{super edge} (bottom), which summarizes how they are connected. Each generated bundle stays linked to its underlying triples and source documents.}
  \label{fig:semantic_bundling_diagram}
\end{figure*}

\begin{figure*}[!t]
  \centering
  \includegraphics[width=\linewidth]{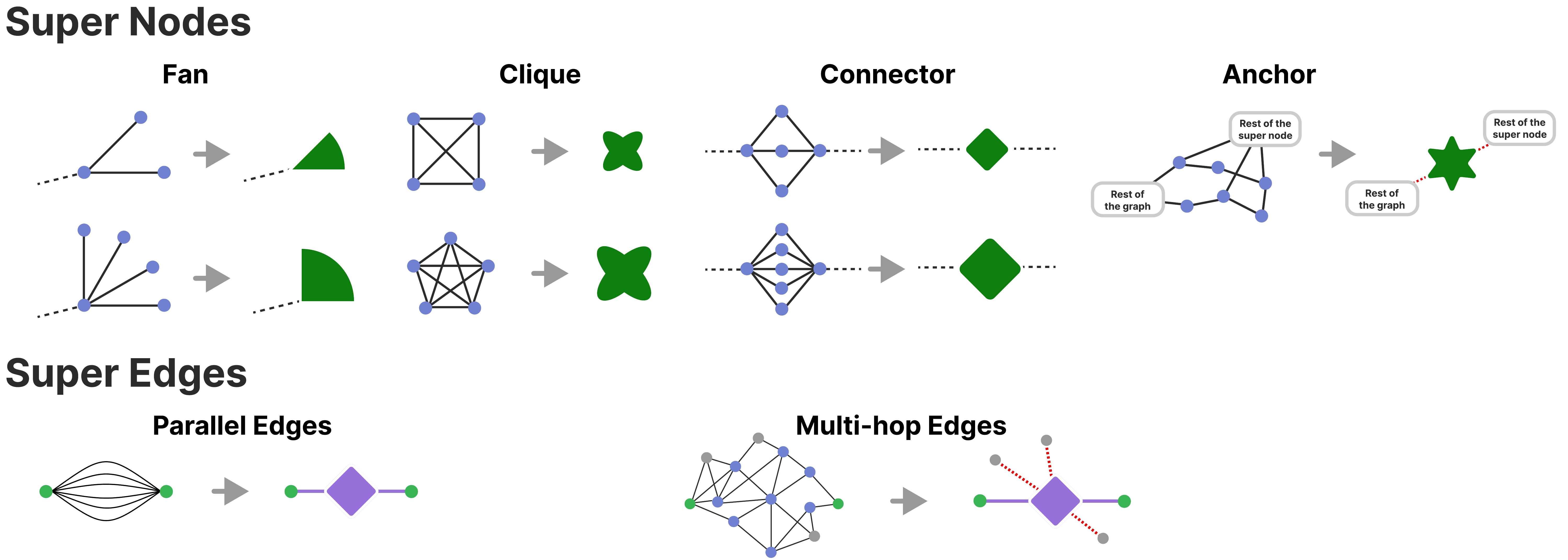}
  \caption{Semantic Bundling uses several glyphs to convey the structure of the consolidated subgraph. \textit{Super nodes} (top) can be composed of up to four glyph types: fan, clique, connector and anchor. The arc length of a fan encodes its number of leaf nodes, while the size of a clique or connector encodes its number of nodes. A \textit{super edge} (bottom) is represented as a diamond. Nodes within anchors or super edges connect to the rest of the graph via red dashed lines.}
  \label{fig:glyph}
\end{figure*}

\subsection{Super Nodes}
\label{sec:sb/super_nodes}

A super node collapses a region of the knowledge graph (i.e., the neighborhood of a single selected entity, or a subgraph the user lassoes) into a single node whose label and natural language summary describe the relationships it contains (Fig.~\ref{fig:semantic_bundling_diagram}).
For example, collapsing the hairball around a company entity yields a single node that summarizes how that company relates to the people and events around it.
Rather than removing the collapsed region, a super node replaces it with a glyph that preserves its structural ``shape''.
Specifically, the super node algorithm detects recurring graph motifs within the region including fans (a hub linked to several leaves), cliques (densely interconnected entities), and connectors (the remaining bridging relationships), leveraging glyph designs inspired by Dunne et al. \cite{dunne2013motif}.
We render these motifs as labeled subcomponents (Fig.~\ref{fig:glyph}).
The result disentangles hairballs in the KG while remaining recognizable enough to act as a navigational landmark for the users' place in the larger graph.

\medskip
\noindent\textbf{Goals.  }
We design super nodes to meet three goals:

\begin{enumerate}
    \item[G1] \textbf{Summarize a region in place} so users can read what a dense subgraph is about without leaving the visual or reading the underlying documents (C1).
    \item[G2] \textbf{Preserve recognizable structure} by encoding the region's motifs as a glyph (C1).
    \item[G3] \textbf{Let users control the granularity of the collapsed super node} by summarizing the region as a whole, by a topic of interest, into a chosen number of groups, or into automatically discovered groups (C1).
\end{enumerate}

\medskip
\noindent\textbf{Algorithm.  }
Given a list of KG triples, we generate super nodes as follows.
Full algorithmic details and LLM prompts can be found in the supplemental materials.

\begin{enumerate}
    \item \textbf{Group the triples} according to a user-specified granularity (G3). Users can summarize all triples as a single cluster, or embed each triple as a text string and cluster the embeddings: by ranking triples against a topic of interest with a cross-encoder and keeping the most similar; into a chosen number $n$ of groups, using agglomerative clustering; or into an automatically chosen number of groups, using HDBSCAN.
    \item \textbf{Label and summarize each group.} For each group of triples, we prompt an LLM to generate a concise topic label and a natural language summary of the relationships, constrained to describe only the relationships present in the group (G1).
    \item \textbf{Link the summary to its triples.} We split the summary into sentences and link each triple to the sentence it best supports using the cross-encoder, so users can trace any part of the summary back to the underlying relationships.
    \item \textbf{Detect and label motifs.} We detect fans, cliques, and connectors within each group (Fig.~\ref{fig:glyph}), then label every motif and every connection that leaves the super node in a single LLM call, producing the labeled glyph (G2).
\end{enumerate}

\subsection{Super Edges}
\label{sec:sb/super_edges}
A super edge summarizes the connection between two entities (Fig.~\ref{fig:semantic_bundling_diagram}).
When the entities are directly connected by several parallel relationships, the super edge collapses them into a single labeled relationship.
When the entities are distant, the super edge finds the multi-hop paths between them and summarizes how they are connected, optionally separating the distinct ways they are linked.
The label of a super edge is itself a KG triple (subject, predicate, object) so that a summarized connection reads as a single relationship the user can reason about in graph terms (C2, C3).
Similar to super nodes, we render the result motif depending on the underlying topology of the KG (Fig.~\ref{fig:glyph}).

\medskip
\noindent\textbf{Goals.  }
We design super edges to meet two goals:

\begin{enumerate}
    \item[G4] \textbf{Summarize a connection as a single interpretable relationship}, whether it comprises many parallel edges between adjacent entities or a long multi-hop path between distant ones (C2, C3).
    \item[G5] \textbf{Surface and contrast the distinct ways two entities are connected} when many paths exist. For example, a user can group paths by shortest path, by the most distinct paths, by paths through specified intermediaries, by topic, or by automatically discovered clusters of paths. This lets users compare alternative reasoning chains that link two distant entities (C3).
\end{enumerate}

\medskip
\noindent\textbf{Algorithm.  }
Given two selected entities, we generate super edges as follows.
Full algorithmic details and LLM prompts can be found in the supplemental materials.

\begin{enumerate}
    \item \textbf{Gather the connecting relationships.} For directly connected entities, we take their parallel edges. For distant entities, we enumerate the simple paths between them, bounded by a maximum length and an optional direction constraint to keep enumeration tractable (Fig.~\ref{fig:glyph}).
    \item \textbf{Group the paths} when more than one exists by a user-specified granularity (G5): into a single shortest path; into the most distinct paths, by selecting those least similar to the rest; into a topic of interest, by cross-encoder ranking; or into an automatically chosen number of groups, using agglomerative clustering with silhouette analysis to select the number of clusters. Users may optionally restrict paths to those passing through specified intermediate entities of interest.
    \item \textbf{Label and summarize each group.} We prompt an LLM to generate a summary triple as the label and a natural language summary of how the entities are connected along those paths (G4).
    \item \textbf{Link the summary to its triples}, similar to super nodes, so every part of the summary is traceable to the relationships and documents beneath it.
\end{enumerate}

\subsection{Grounding and Traceability}
\label{sec:sb/grounding}

We design both operations to keep their generated summaries grounded in the source data, so every generated summary stays traceable to its evidence (C4).
First, we constrain the LLM at every step to summarize only the relationships present in the selected region or paths; in effect, the LLM labels and condenses the graph but does not speculate.
Second, we link each generated summary back to the data it derives from: we associate every triple with the summary sentence it best supports, and each triple already links to its source documents.
For example, a user reading a super edge summary can click any sentence to highlight the exact relationships it came from, then open those relationships' source documents to confirm the claim.
Together, these let users move from any sentence of a summary to the underlying relationships, and from there to the source text.

\section{The Agent{K} System}
\label{sec:system}

\begin{figure*}[!t]
  \centering
  \includegraphics[width=\linewidth]{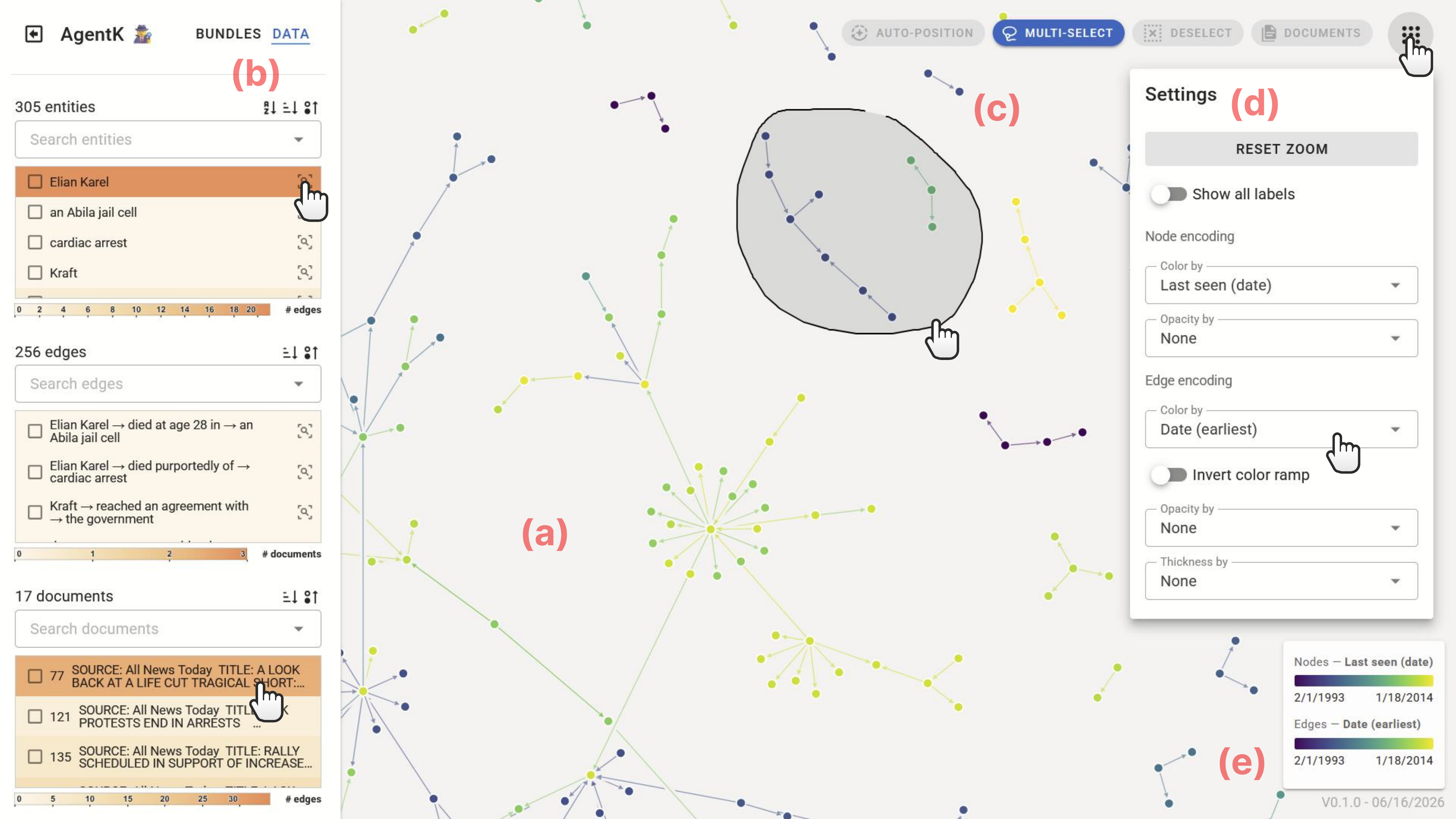}
  \caption{The AgentK main view loaded with the KRONOS dataset~\cite{whiting2014vast}. Users (a) explore the knowledge graph as a force-directed node-link diagram, (b) use the side panel to browse and select the underlying triples and source documents, and (c) select a node, subgraph, or pair of entities to bundle. Users can encode graph metadata (type, date, sentiment, etc.) using color, opacity, and thickness (d), with an accompanying legend automatically generated (e).}
  \label{fig:agentk_main}
\end{figure*}

To realize Semantic Bundling, we built AgentK, an open-source proof-of-concept visual analytics system for analyzing text documents as KGs.
AgentK takes a collection of documents, represents them as a KG, and lets a user transform that graph into super nodes and super edges through direct interaction.
In this section we describe how AgentK visualizes documents as a KG (Sect.~\ref{sec:agentk/visualizing}), maps user interactions to Semantic Bundling operations (Sect.~\ref{sec:agentk/transforming}), and visualizes the resulting bundles for inspection (Sect.~\ref{sec:agentk/inspecting}).

\subsection{Visualizing Text Documents as a Knowledge Graph}
\label{sec:agentk/visualizing}

The central workflow of AgentK involves uploading raw text documents, converting them into a KG, and providing a node-link visualization to explore the generated KG, with features to connect the KG back to the source documents it came from.

\medskip
\noindent\textbf{Uploading and extracting a KG.  }
AgentK can construct a KG directly from a set of raw text documents.
When a user uploads documents, AgentK extracts a graph using an LLM: for each document, it extracts the checkable claims as $\langle$\textit{subject, predicate, object}$\rangle$ triples.
At this stage, the extraction pipeline resolves pronouns to the entities they refer to and classifies each entity and relationship with a type and sentiment, together with document-level metadata such as date and source.
The pipeline then reduces similar triples through semantic deduplication \cite{abbas2023semdedup} and assembles the result into a single graph, propagating the extracted metadata onto its nodes and edges.
Because the LLM defines this ontology at extraction time rather than fixing it in advance, entities and relationships use open, human-readable vocabulary; we explore the consequences of this design in our usage scenarios (Sect.~\ref{sec:usage_scenarios}).
In addition to raw text documents, users may also upload an existing knowledge graph; the extraction step is a convenience for the common case where only the source documents are at hand, much as Jigsaw provides a built-in entity extractor while still accepting pre-extracted data~\cite{Stasko:2007:Jigsaw}.

\begin{figure*}[!t]
  \centering
  \includegraphics[width=\linewidth]{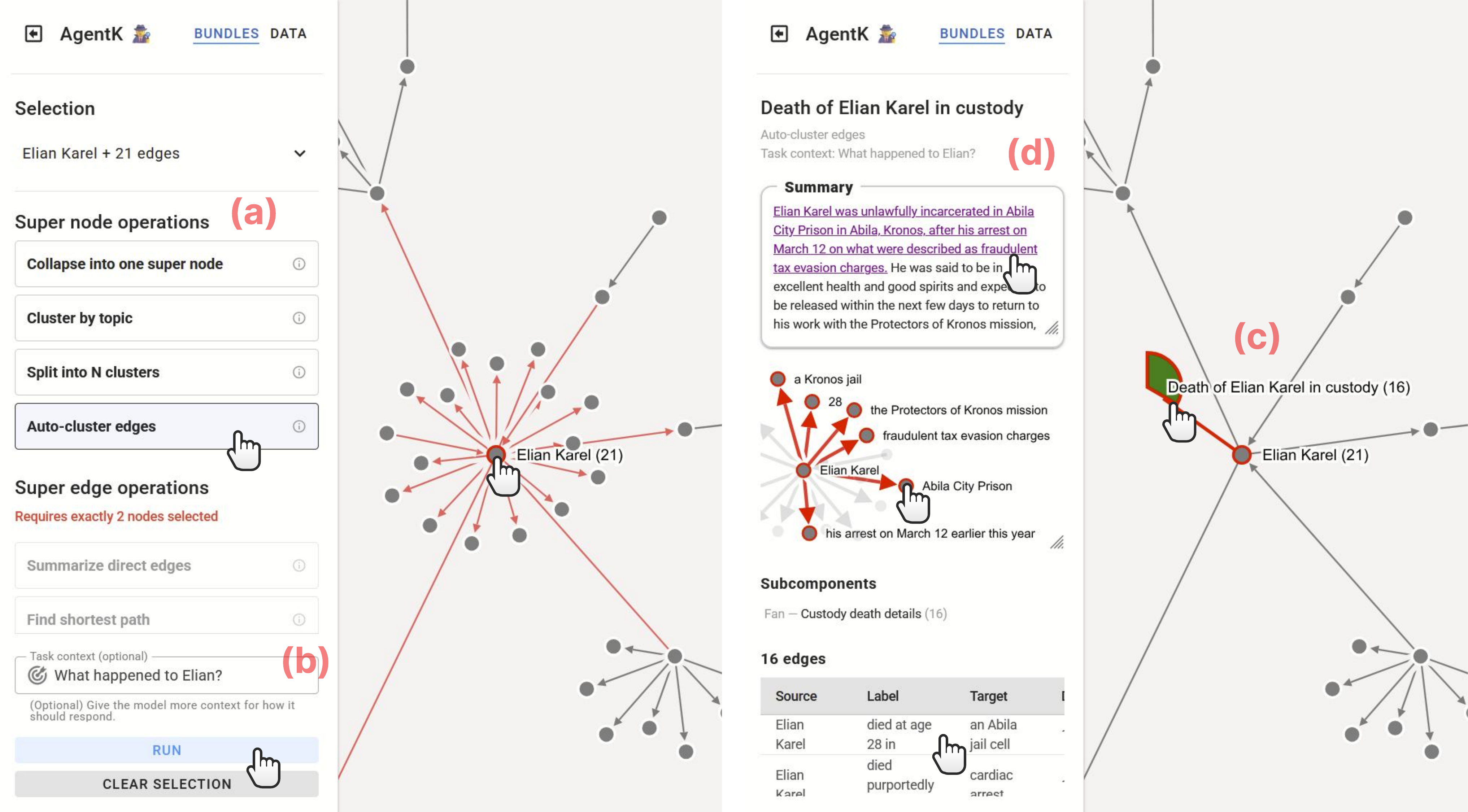}
  \caption{The AgentK results view loaded with the KRONOS dataset~\cite{whiting2014vast}. Selecting nodes/edges in the main view opens the Bundles tab (a), which exposes options for applying a bundling operation based on the user's selection, with optional task context free-text input (b). Semantic Bundles are automatically visualized as glyphs in the graph (c). Selecting a Semantic Bundle reveals a results panel with details on demand (d).}
  \label{fig:agentk_results}
\end{figure*}

\medskip
\noindent\textbf{Views.  }
AgentK presents the KG as a force-directed node-link diagram in the main view (Fig.~\ref{fig:agentk_main}a) with two side panels: a \textit{Data} tab and a \textit{Bundles} tab.
The Data tab (Fig.~\ref{fig:agentk_main}b) provides sortable, searchable tables of the graph's entities, relationships, and source documents, and supports brush-and-link interaction, inspired by Jigsaw's List View feature \cite{Stasko:2007:Jigsaw}.
Selecting a row highlights the corresponding node or edge in the graph.
Selecting a document opens a document viewer that shows its full text alongside the triples extracted from it, so the user can always return to the source evidence (C4).
The Bundles tab, described in Sect.~\ref{sec:agentk/transforming}, is where users configure Semantic Bundling operations and inspect their results.

\medskip
\noindent\textbf{Controls, settings, and encodings.  }
Users navigate the graph with mouse pan, zoom, and drag controls, and select regions of interest by clicking individual nodes and edges or by lassoing a subgraph (Fig.~\ref{fig:agentk_main}c).
AgentK provides standard graph encodings based on metadata extracted during construction that can be applied to the visualization (Fig.~\ref{fig:agentk_main}d): node color and opacity can encode an entity's type, dominant sentiment, mention count, document count, or first/last appearance; edge color, opacity, and width can encode a relationship's type, sentiment, date, or frequency.
AgentK shows a legend for any active encodings (Fig.~\ref{fig:agentk_main}e).
For example, coloring nodes by mention count surfaces the busiest entities at a glance, a natural starting point for bundling.
These encodings give users an overview of meaning and activity in their document corpus (C1, C2).

\subsection{Transforming Knowledge Graphs using Semantic Bundles}
\label{sec:agentk/transforming}

AgentK's core functionality brings Semantic Bundling into an interactive visual analytics workflow.
As users directly manipulate the KG visualization, they can generate Semantic Bundles to disentangle hairballs and draw out connections between disparate nodes, aiding their analysis.

\medskip
\noindent\textbf{Mapping user selection to bundling operations.  }
AgentK infers which Semantic Bundling operation a user wants from the structure of their selection, bridging low-level graph interactions and high-level analytic goals.
Selecting a single node, or lassoing a subgraph, produces a super node that summarizes the selected region.
Selecting two directly connected entities produces a super edge over their parallel relationships, while selecting two disconnected entities produces a super edge over the multi-hop paths between them.
Compared to menu-based invocation, our mapping bridges a gap in direct manipulation approaches to KG simplification for visual analytics.

\medskip
\noindent\textbf{Operations and settings.  }
Once a selection is made, the Bundles tab exposes operation settings (Fig.~\ref{fig:agentk_results}a).
For a super node, the user chooses the granularity of the collapse: the whole region, a topic of interest entered as text, a chosen number of groups, or automatically discovered groups.
For a super edge between distant entities, the user chooses how to gather and group paths (shortest, most distinct, through specified intermediaries, by topic, or automatically) and bounds the search with a maximum path length and a directionality toggle.
We also provide a global task context field as a free-text input (Fig.~\ref{fig:agentk_results}b).
This lets the user state the question or goal driving their analysis, which AgentK then supplies to every operation as context for the LLM.
The context helps steer the LLM's labels and summaries toward what the user cares about.

\subsection{Inspecting Semantic Bundles}
\label{sec:agentk/inspecting}

AgentK provides details-on-demand visualizations that trace Semantic Bundles back to the original graph structure and the underlying triples with their source documents.

\medskip
\noindent\textbf{Visualizing bundles directly in the graph.  }
AgentK renders super nodes and super edges directly in the node-link diagram (Fig.~\ref{fig:agentk_results}c), allowing a bundle to take the place of the structure it summarizes.
A super node appears as an anchor glyph from which its motifs radiate as distinct labeled subcomponents, preserving the recognizable shape of the collapsed region (G2).
A super edge appears as a connector glyph between its two entities, with the contributing paths drawn as bundled edges so the most traveled connections remain visible.
AgentK draws relationships that leave a bundle as dashed links, visually distinguishing the generated structures from the original graph and signaling that the bundle still connects to the larger graph.

\medskip
\noindent\textbf{Details on demand.  }
Selecting a bundle opens its result in the Bundles side panel (Fig.~\ref{fig:agentk_results}d).
For each Semantic Bundle, the panel displays the generated label, the natural language summary, a mini graph of the bundled relationships, and a table of the underlying triples.
The summary is interactive: hovering or clicking a sentence highlights the triples it links to in both the mini graph and the table.
Selecting a triple opens its source documents in the document viewer.
A user can therefore read a fluent summary, confirm any claim against the relationships and documents beneath it, and optionally delete a bundle to restore the original graph structure.

\subsection{Implementation}
\label{sec:agentk/implementation}
We built AgentK's interface with Vue and used D3 for the force-directed graph.
The frontend communicates with a Python Flask backend that runs the graph algorithms (path finding, motif and clique detection, and clustering) and orchestrates the LLM operations.
We use OpenAI's GPT-5.2 for knowledge graph extraction, labeling, and summarization, embed triples with the \texttt{all-MiniLM-L6-v2} sentence encoder, rank and link text with the \texttt{stsb-distilroberta-base} cross-encoder, and deduplicate triples with SemHash.
Graph operations use NetworkX.
AgentK is open-source and available at [blinded for review].

\section{Usage Scenarios}
\label{sec:usage_scenarios}

To demonstrate how AgentK facilitates text document analysis, we first present insights from the IMDb movie reviews dataset~\cite{zm1y-b270-20}. We then situate these capabilities within a real-world intelligence analysis scenario through a fictitious case study with KRONOS dataset~\cite{whiting2014vast}.

\subsection{Insights Discovery: IMDb Movie Reviews Dataset}

We construct a knowledge graph from the top 400 ``most helpful" reviews for the movie \textit{Interstellar} (2014) from the classic IMDb reviews dataset~\cite{zm1y-b270-20}. The following subsections demonstrate how AgentK addresses the five core design goals (\textbf{G1}-\textbf{G5}) described in Sections~\ref{sec:sb/super_nodes} and \ref{sec:sb/super_edges}.

\subsubsection{G1: Summarize a Region in Place}

\begin{figure}[!t]
  \includegraphics[width=\linewidth]{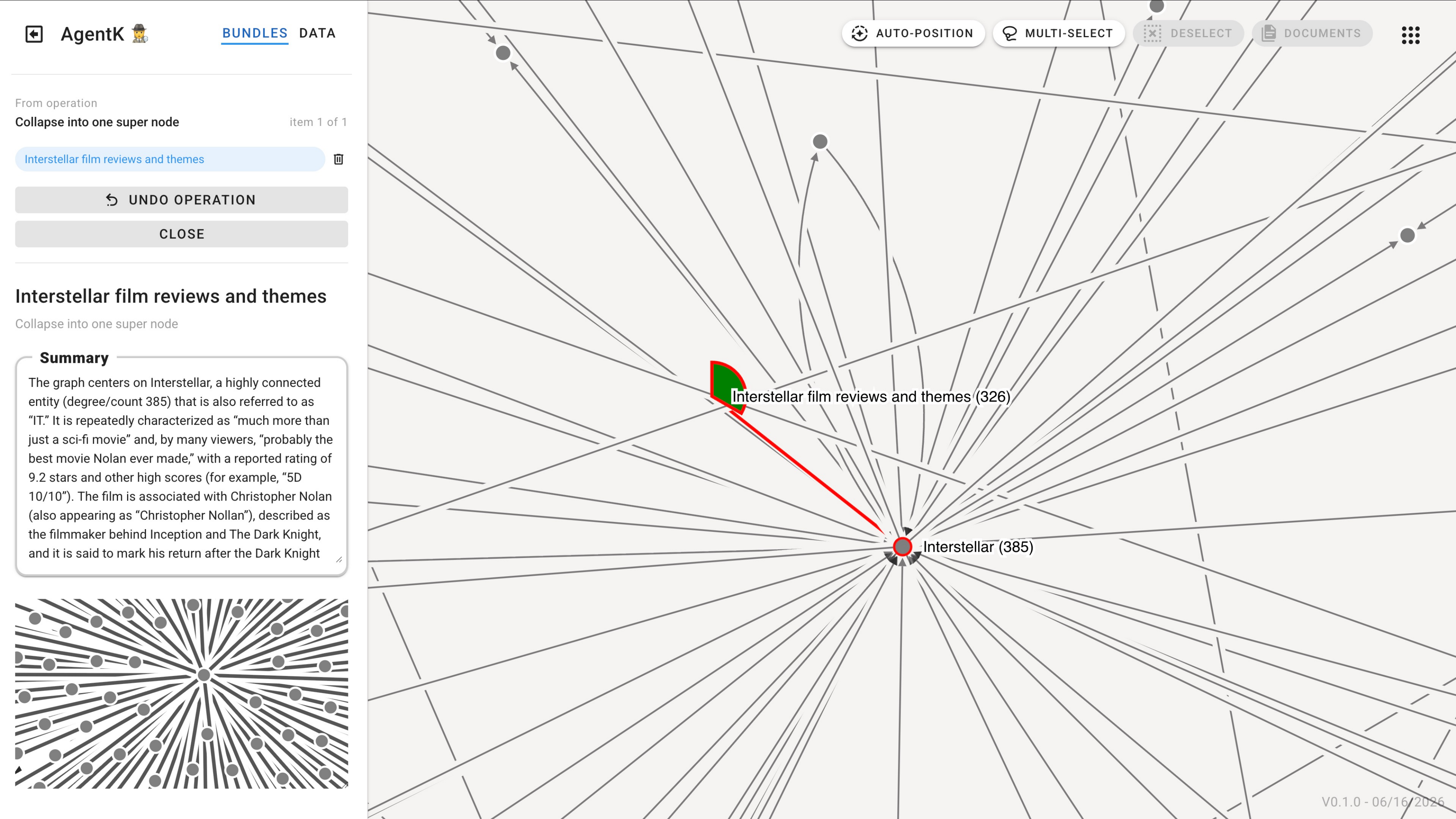}
  \caption{The dense hairball surrounding the node ``Interstellar" is consolidated into a super node labeled ``Interstellar film reviews and themes'', reducing visual clutter while preserving entity-level context.}
  \label{fig:imdb_interstellar_super_node}
\end{figure}

The node ``Interstellar" serves as a high-degree hub, connected to a dense hairball of edges. Rather than exposing users to this visual complexity, AgentK's super node feature enables them to consolidate the surrounding structure into a single, semantically meaningful node (Fig.~\ref{fig:imdb_interstellar_super_node}). The hairball is replaced by a super node labeled ``Interstellar film reviews and themes'', while entity-level details such as production background and user ratings remain accessible.

\subsubsection{G2: Preserve Recognizable Structure}

\begin{figure}[!t]
  \includegraphics[width=\linewidth]{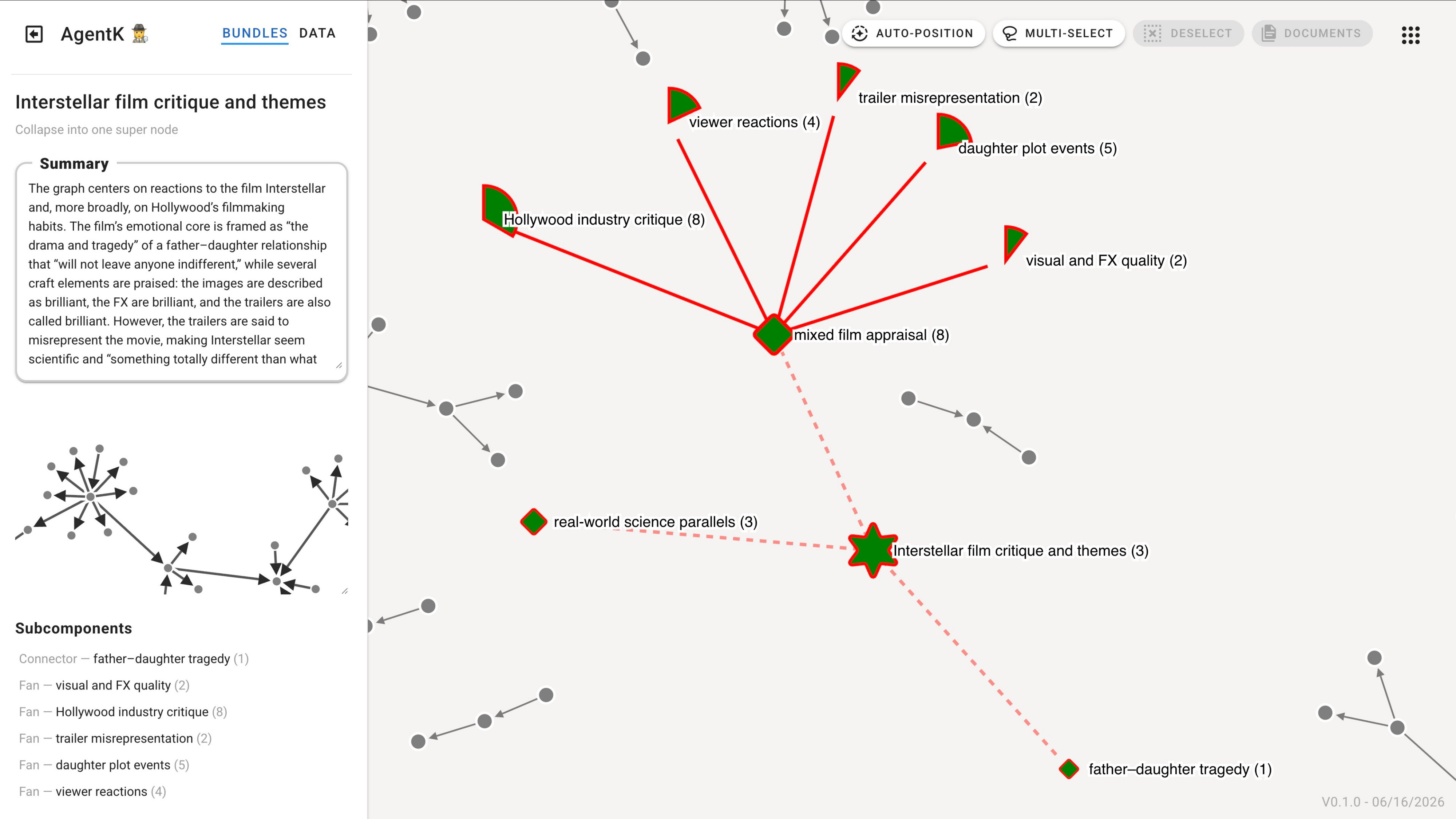}
  \caption{User-selected subgraph is abstracted into a super node, preserving both high-level thematic categories and lower-level structural details.}
  \label{fig:imdb_subgraph_labeling}
\end{figure}

AgentK's super node feature also enables users to label and abstract complex subgraphs. When a user identifies a densely connected subgraph, such as a cluster of nodes discussing the quality of trailer, they can lasso the selection 
and convert it into a single super node titled ``Interstellar film critique and themes" (Fig.~\ref{fig:imdb_subgraph_labeling}). Unlike traditional node collapsing, the super node motif preserves the structural shape of the original subgraph while independently labeling each constituent part. This allows users to shift between high-level thematic overviews and lower-level structural details without discarding either.

\subsubsection{G3: Let the User Control the Granularity of a Collapse}

\begin{figure}[!t]
  \includegraphics[width=\linewidth]{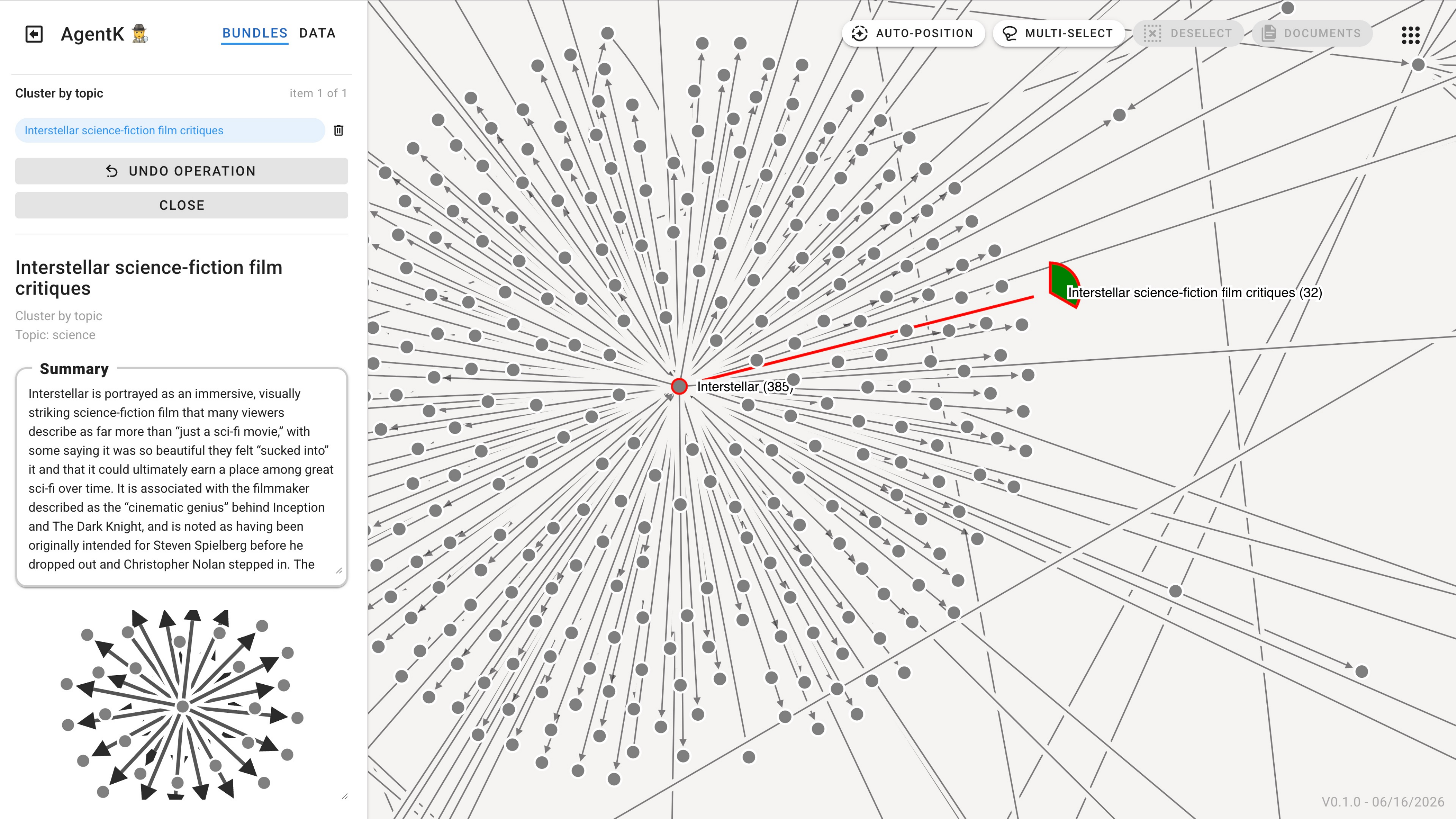}
  \caption{Nodes related to the focal topic ``science'' are selectively clustered and summarized into a single super node, illustrating user-controlled granularity.}
  \label{fig:imdb_interstellar_science}
\end{figure}

AgentK enables users to synthesize relationships at varying levels of granularity. At the coarsest level, the user can collapse the entire hairball surrounding the ``Interstellar" node into a single super node, as described above (Fig.~\ref{fig:imdb_interstellar_super_node}). At a finer level, the user may supply a focal topic such as ``science" to filter the surrounding structure, directing AgentK to perform topical relationship inference by clustering and summarizing only those nodes relevant to that topic. As illustrated in Fig.~\ref{fig:imdb_interstellar_science}, this produces a compact super node titled ``Interstellar science-fiction film critiques.'' This granularity control enables users to investigate specific thematic dimensions of an entity without losing the broader graph context.

\subsubsection{G4: Summarize a Connection as a Single Interpretable Relationship}

\begin{figure}[!t]
  \includegraphics[width=\linewidth]{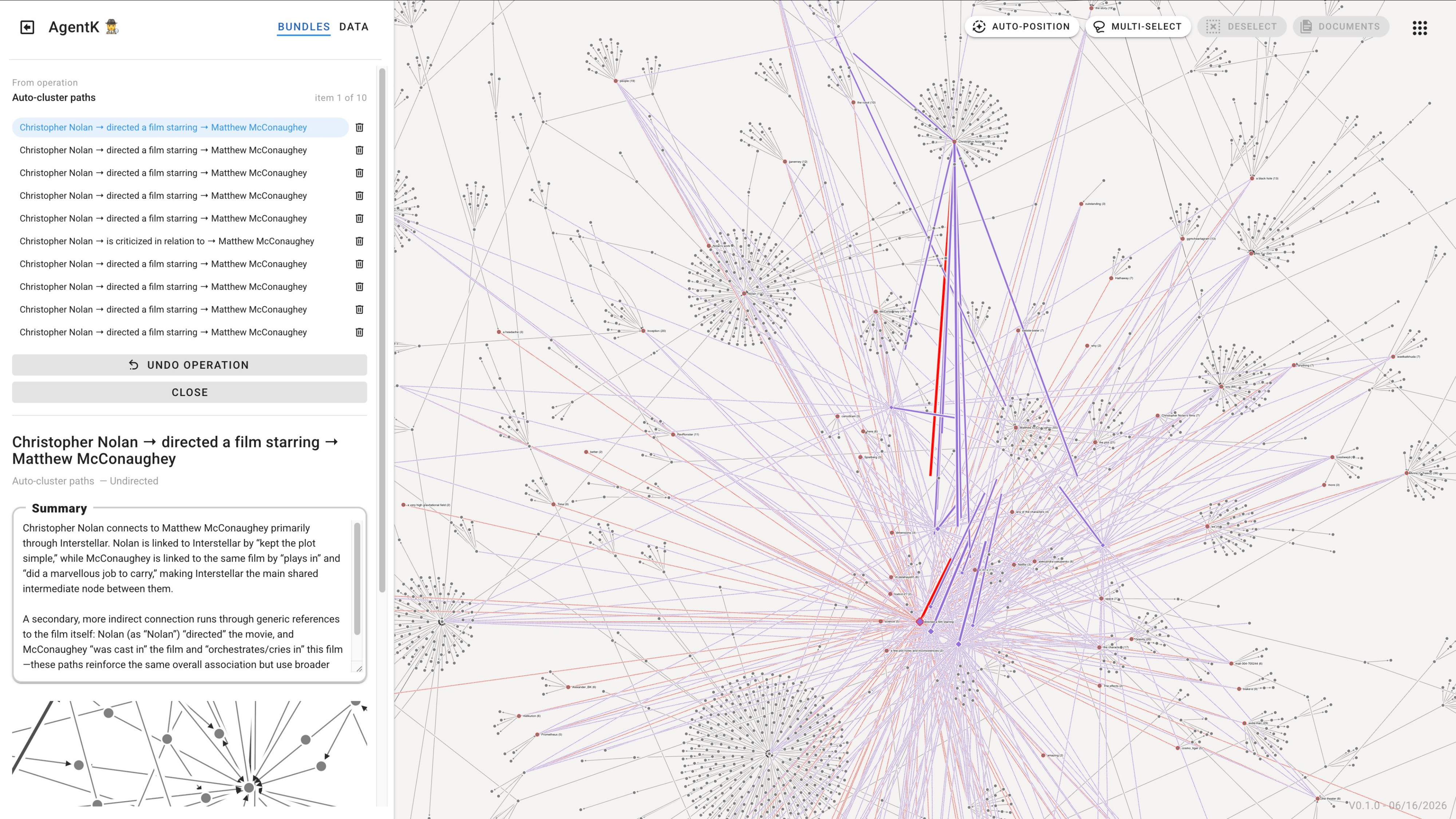}
  \caption{A logical multi-hop relationship between Nolan and McConaughey is discovered and rendered as a super edge through path-finding algorithms.}
  \label{fig:imdb_nolan_and_mcconaughey}
\end{figure}

AgentK supports the discovery of connections between distant, disconnected entities through its super edge feature in conjunction with path-finding algorithms. For example, while no direct edge initially linked director Christopher Nolan to actor Matthew McConaughey, AgentK infers and surfaces the logical connection ``Christopher Nolan $\rightarrow$ directed a film starring $\rightarrow$ Matthew McConaughey" (Fig.~\ref{fig:imdb_nolan_and_mcconaughey}). This capability allows users to uncover higher-order relationships spanning multiple hops across entities.

\subsubsection{G5: Surface and Contrast the Distinct Ways Two Entities are Connected}

\begin{figure}[!t]
  \hfill
  \subfloat[]{\includegraphics[width=\linewidth]{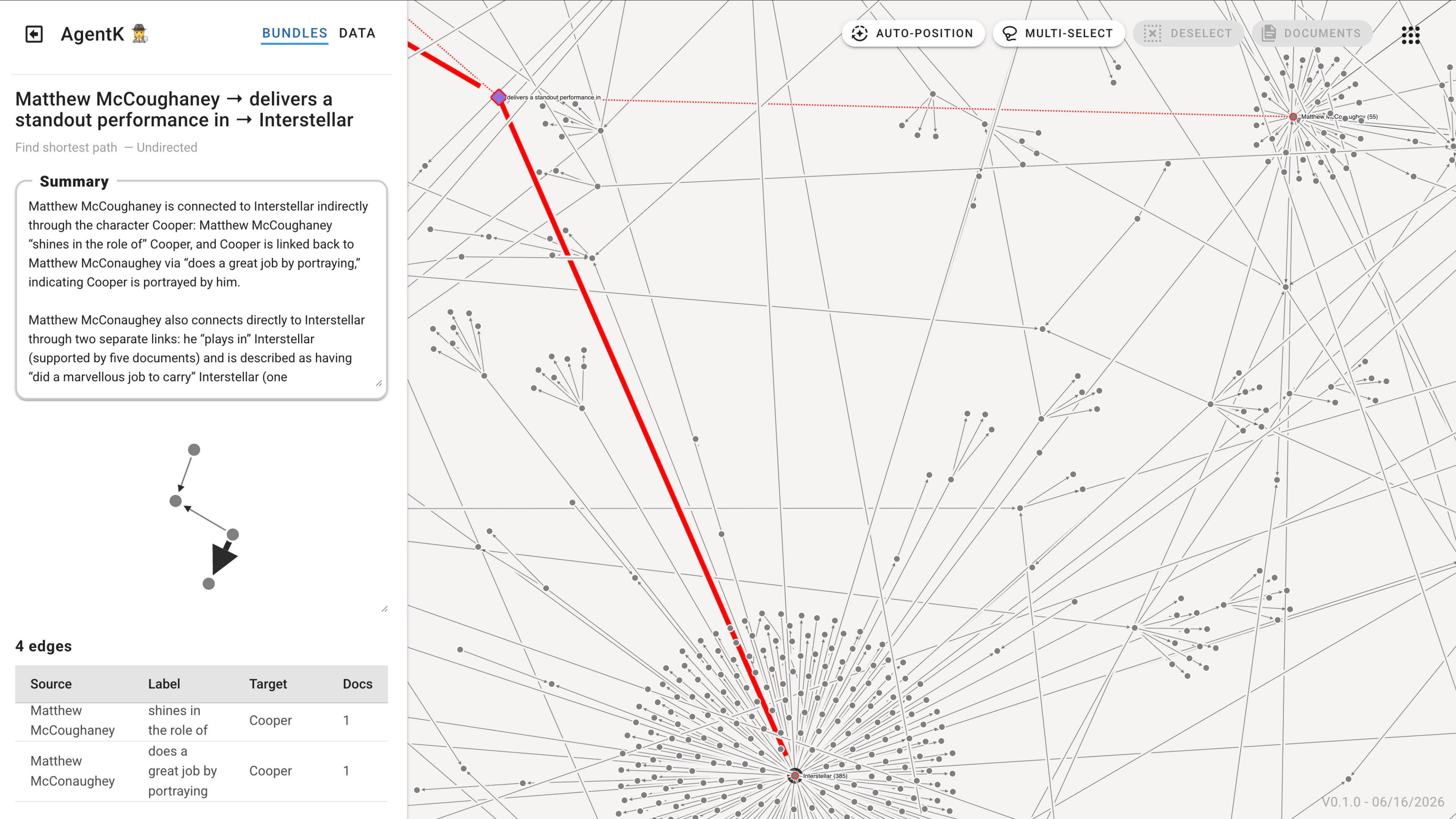}}
  \hfill
  \subfloat[]{\includegraphics[width=\linewidth]{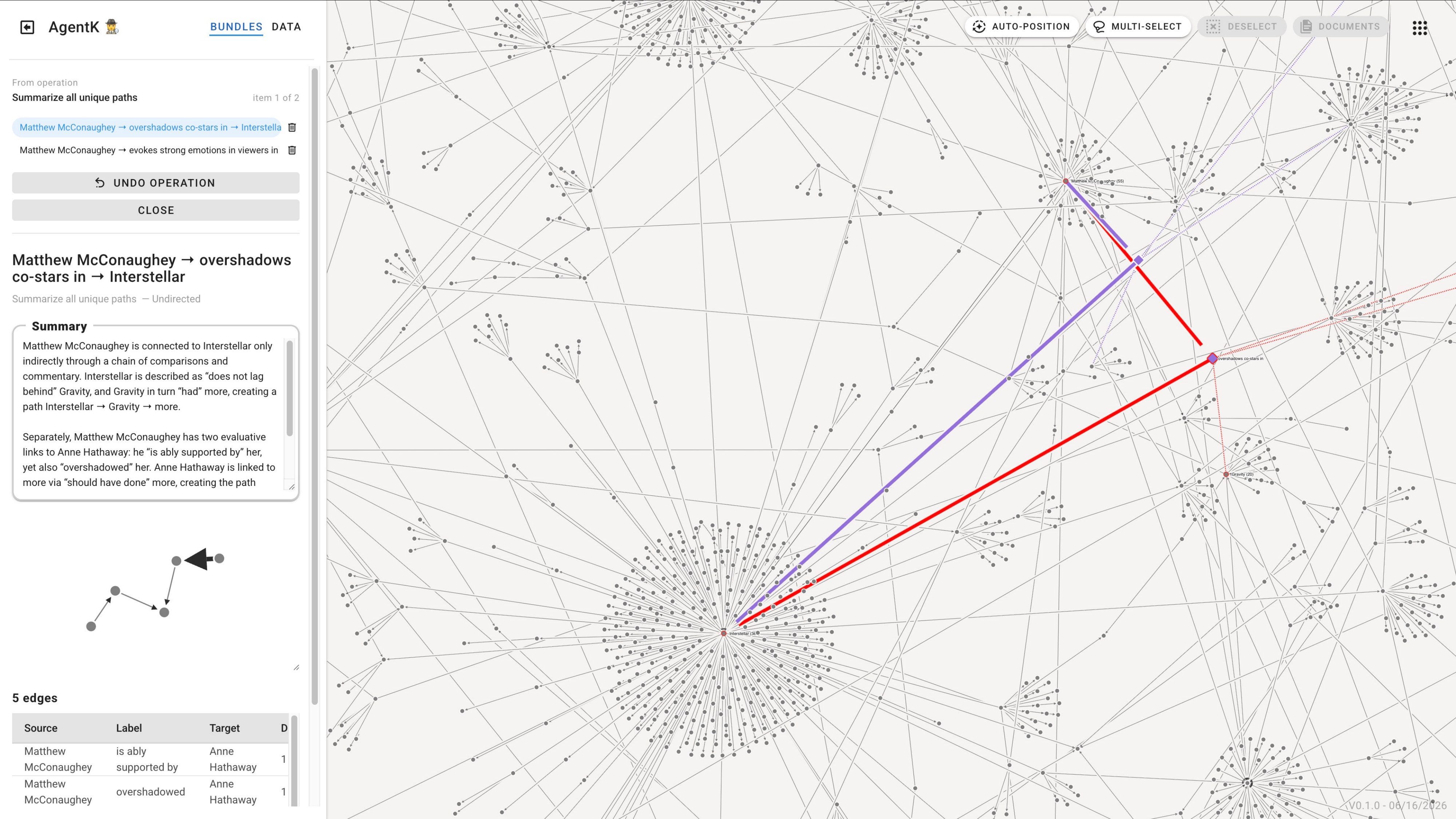}}
  \caption{Super edges generated between Matthew McConaughey and Interstellar using (a) the shortest path and (b) all unique paths, illustrating how different path-finding strategies surface distinct relationship semantics.}
  \label{fig:imdb_matt_interstellar_diff_paths}
\end{figure}

When multiple paths exist between two nodes, AgentK offers several path-finding options that yield qualitatively distinct super edges. Using the shortest path option, AgentK surfaces a predominantly positive audience assessment of McConaughey's performance, representing it as the relationship ``Matthew McConaughey $\rightarrow$ delivers a standout performance in $\rightarrow$ Interstellar" (Fig.~\ref{fig:imdb_matt_interstellar_diff_paths}(a)). When the ``all unique paths'' option is applied, AgentK additionally uncovers comparative sentiment, revealing that audiences not only praised McConaughey's performance but also contrasted it favorably against co-star Anne Hathaway, represented as ``Matthew McConaughey $\rightarrow$ overshadows co-stars in $\rightarrow$ Interstellar" (Fig.~\ref{fig:imdb_matt_interstellar_diff_paths}(b)). These results demonstrate how different path-finding strategies can surface qualitatively distinct relationship semantics from the same underlying graph.

\subsection{Intelligence Analysis: KRONOS Dataset}

To further evaluate AgentK in a practical intelligence context, we apply it to the KRONOS dataset. This investigation follows Betty, a veteran detective tasked with reopening a cold kidnapping case from 2014. Betty aims to uncover latent links between peripheral suspects and the abduction. Faced with a corpus of over 800 news articles, Betty utilizes AgentK to streamline her foraging and synthesis efforts.

\subsubsection{Identifying and Summarizing Salient Entities}

\begin{figure}[!t]
  \centering
  \includegraphics[width=\linewidth]{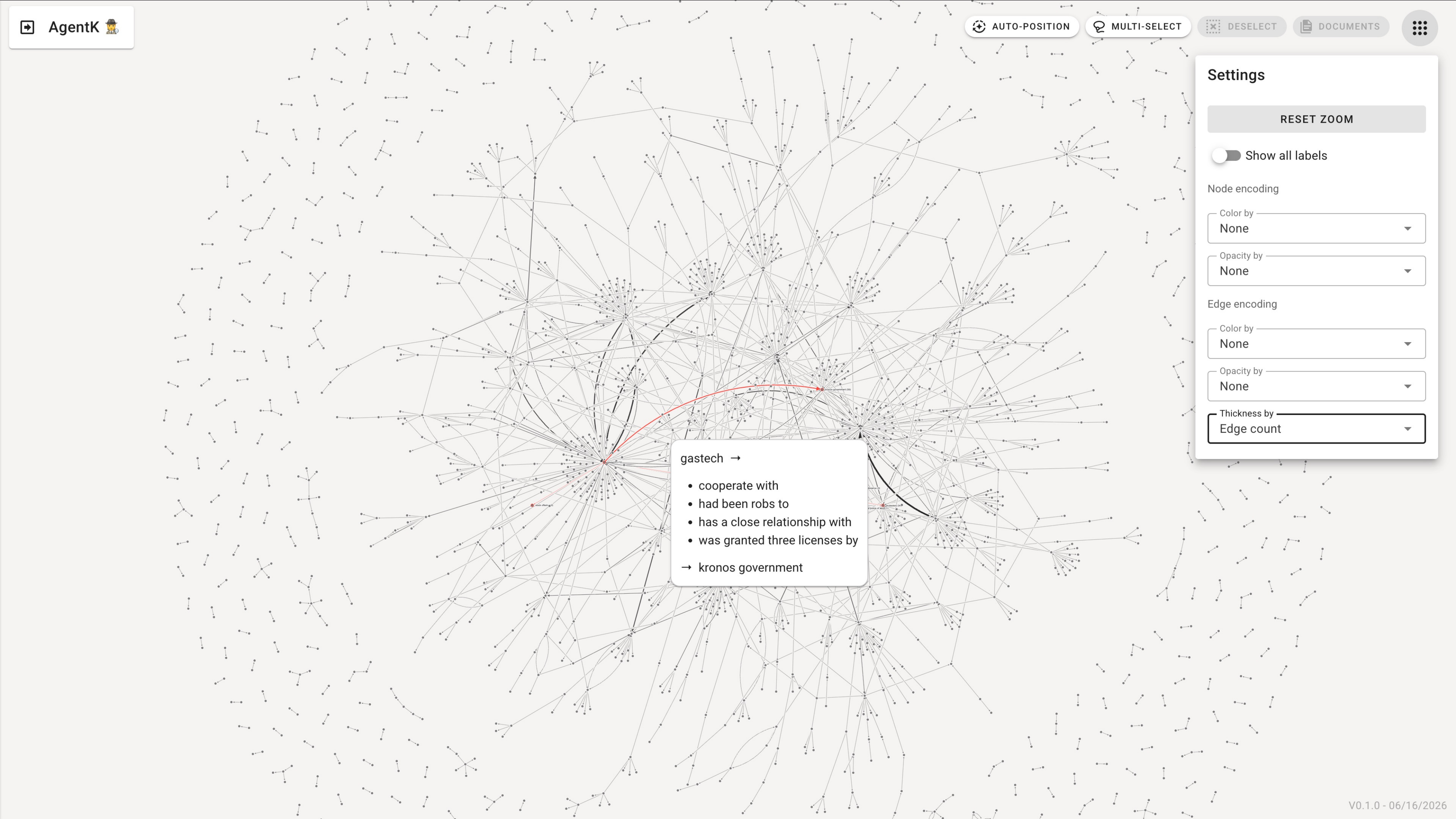}
  \caption{The initial overview of the KRONOS dataset in AgentK. The investigator identifies a dense hairball cluster around the node ``gastech" after enabling edge thickness encoding on edge counts and hovers over an edge to inspect relational details.}
  \label{fig:kronos_overview}
\end{figure}

As Betty begins her investigation, she lacks prior knowledge of the primary actors. The graph visualization in AgentK aids her initial foraging.
Upon loading the graph, Betty quickly identifies several dense hairball clusters and prominent edges by enabling edge thickness encoding by edge counts (Fig.~\ref{fig:kronos_overview}).

\begin{figure}[!t]
  \hfill
  \subfloat[]{\includegraphics[width=\linewidth]{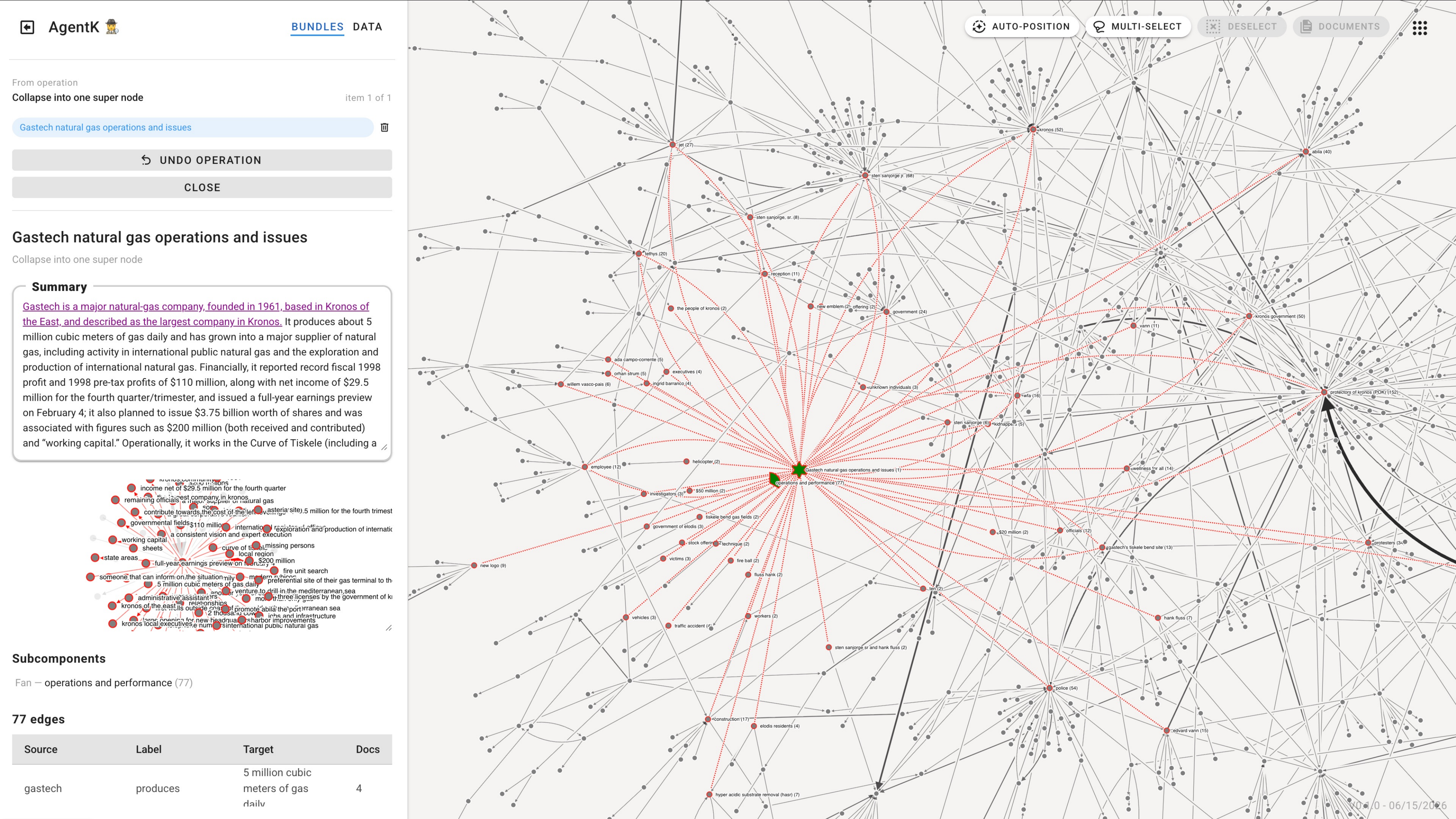}}
  \hfill
  \subfloat[]{\includegraphics[width=\linewidth]{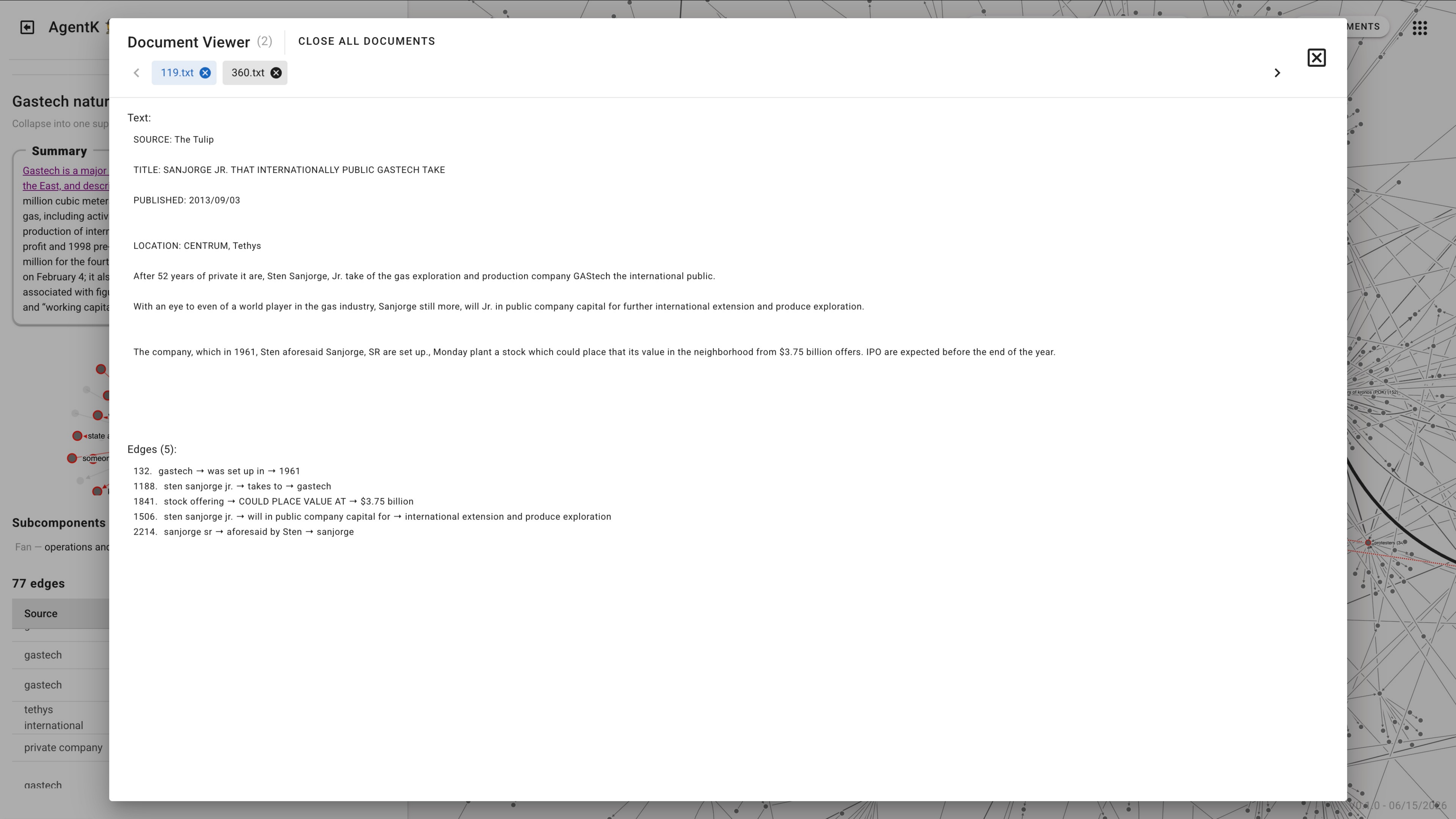}}
  \caption{Betty's super node workflow: (a) Inspecting the semantic summary of a super node aggregated by a hairball cluster (b) Using the document viewer for evidence validation.}
  \label{fig:kronos_generate_supernode}
\end{figure}

She selects a hairball cluster surrounding the node ``gastech" and generates a super node to summarize the entity's context (Fig.~\ref{fig:kronos_generate_supernode}).
The generated super node summary reveals that ``gastech" is a major natural gas company with significant social and political influence in the country of Kronos.
To validate this summary, Betty hovers over sentences in the summary to highlight the node-link relationships in a graph thumbnail.
She further drills down into the relationship table beneath the thumbnail to access the source news articles. Through this iterative process, Betty identifies and contextualizes other key entities, including the ``Protectors of Kronos" (POK), Elian Karel, and Sten Sanjorge Jr.

\subsubsection{Relational Path Analysis}

\begin{figure}[!t]
  \hfill
  \subfloat[]{\includegraphics[width=\linewidth]{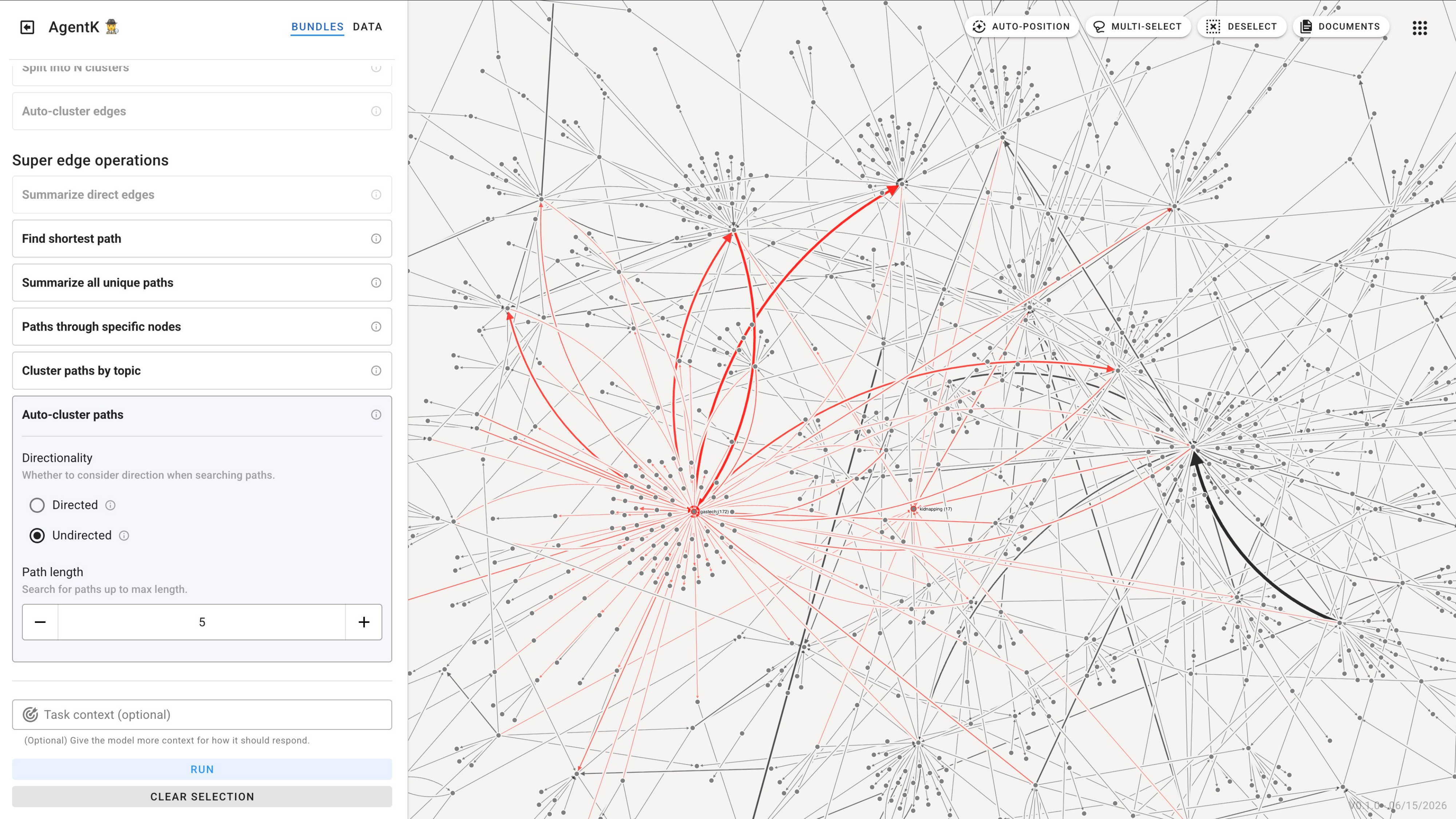}}
  \hfill
  \subfloat[]{\includegraphics[width=\linewidth]{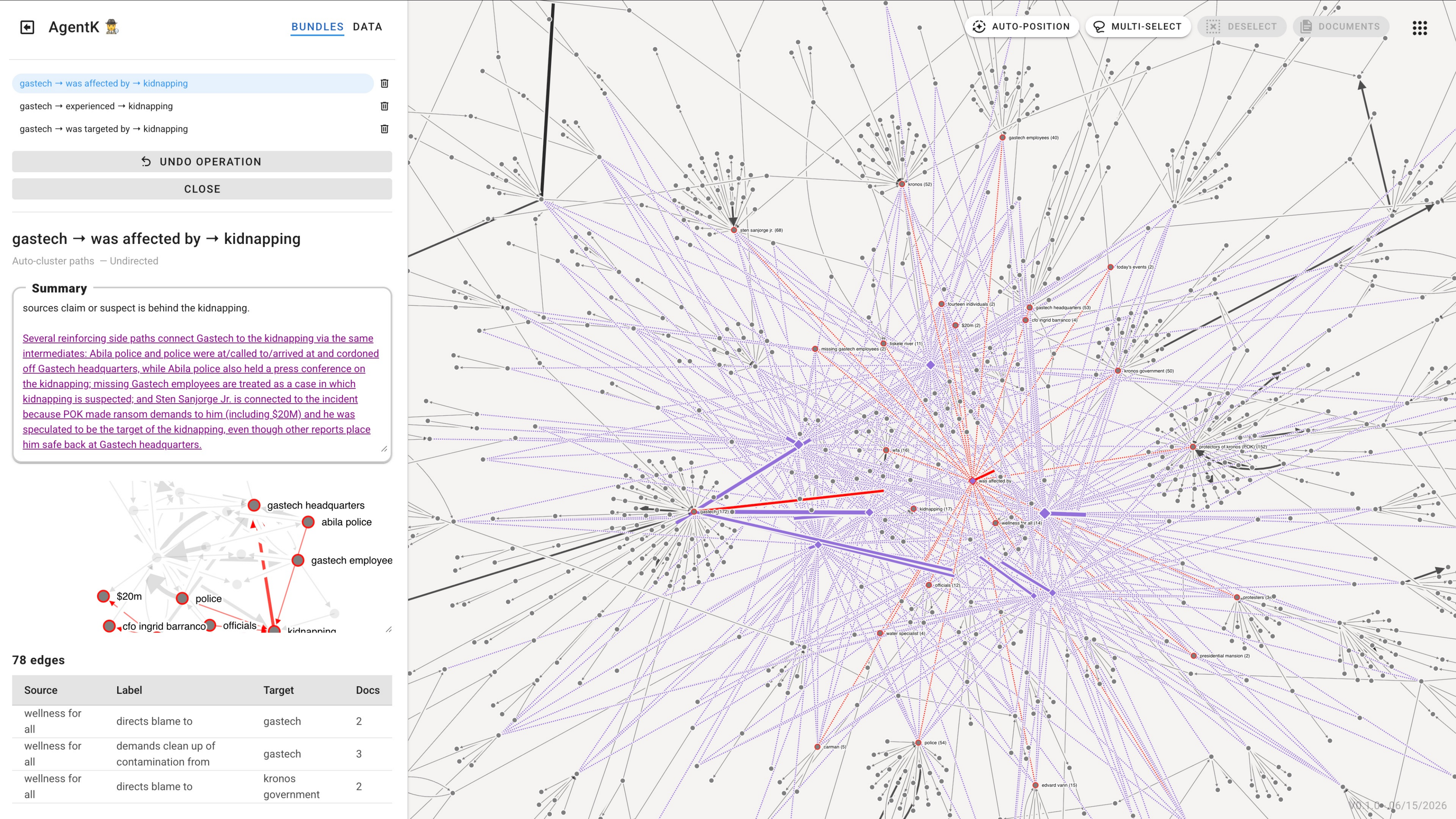}}
  \caption{Betty's super edge workflow: (a) Selecting two nodes to aggregate all paths into super edges (b) Analyzing the summaries of these multi-hop connections.}
  \label{fig:kronos_generate_superedge}
\end{figure}

While the super nodes summaries explain connectivity with adjacent nodes, Betty needs to understand the direct and indirect links between these actors and the kidnapping event. 
To analyze the connection between ``gastech" and ``kidnapping", she  selects both nodes and clusters all paths between them into super edges (Fig.~\ref{fig:kronos_generate_superedge}).

The resulting path summaries reveal something unusual regarding Sten Sanjorge Jr., the CEO of GAStech. Sanjorge Jr. was the presumed target of the kidnapping yet remained unharmed, while several high-ranking executives were taken. Another path summary mentions that the GAStech employee Edvard Vann was interrogated by the police due to a name similarity with some known POK members.
By inspecting the original documents, Betty discovers that Sanjorge Jr. was absent from a mandatory board meeting during the exact hours of kidnapping. Based on these insights, she flags Edvard Vann and Sten Sanjorge Jr. as primary persons of interest.

\subsubsection{Validate Context and Discover Intent}

\begin{figure}[!t]
  \hfill
  \subfloat[]{\includegraphics[width=\linewidth]{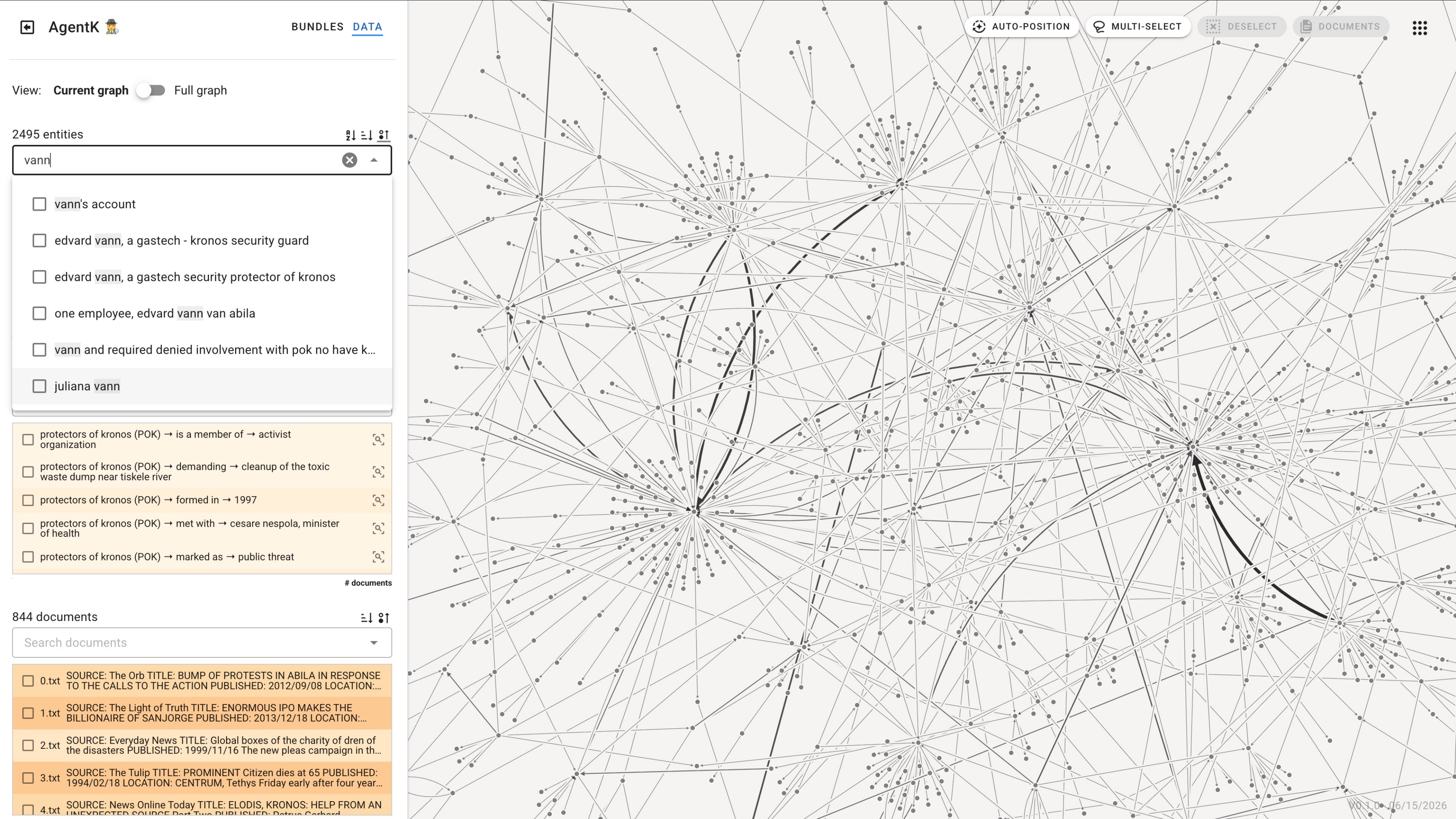}}
  \hfill
  \subfloat[]{\includegraphics[width=\linewidth]{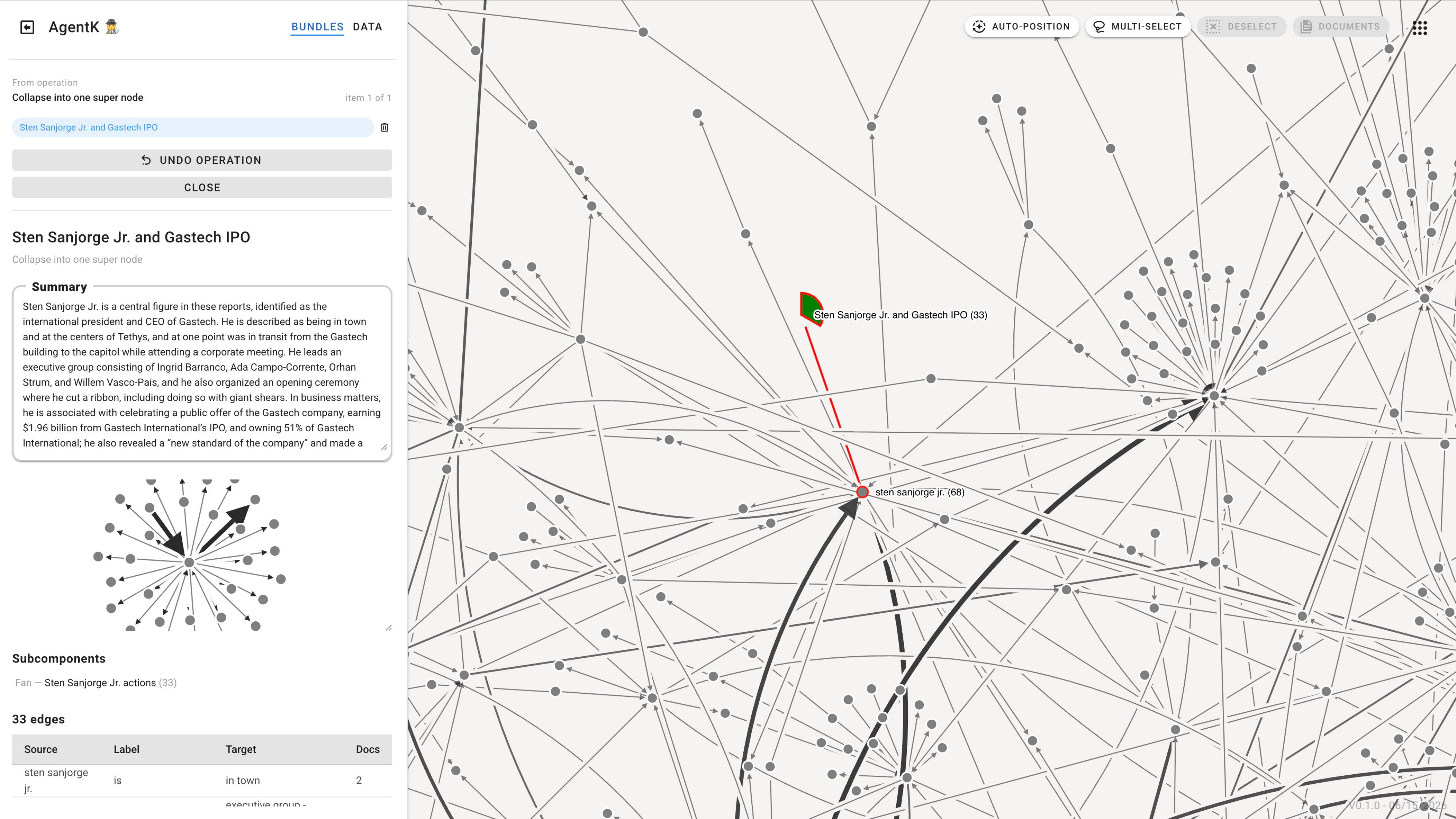}}
  \hfill
  \caption{Betty's intent discovery workflow: (a) Querying keywords in the data viewer (b) Super node clustering performed earlier.}
  \label{fig:kronos_intent_discovery}
\end{figure}

With a list of suspects established, Betty needs a deep dive into the documents to uncover their motives and find more evidence (Fig.~\ref{fig:kronos_intent_discovery}).

She first queries ``vann" in the data viewer and identifies ``Juliana Vann", a young girl whose death from water contamination made her the symbolic face of the POK movement against GAStech.
This gives Edvard Vann, the GAStech security guard, a clear and concrete motive and his internal security knowledge would help facilitate the kidnapping.

Simultaneously, Betty investigates the financial context surrounding Sanjorge Jr. to understand why he might benefit from being away during the attack. She recalls that Sanjorge Jr. earned \$1.96 billion from the GAStech International IPO.
The elimination of his fellow executives would consolidate his leadership and remove potential internal challenges while the company was at peak value.
Betty also notes a discrepancy in the ransom because POK requested only \$20 million, which is just a small fraction of the recent IPO earnings.

With these facts combined, Betty hypothesizes that Edvard Vann may be the operational perpetrator, while Sten Sanjorge Jr. acted as the organizer in the shadows. His absence from the event was likely a coordinated move to ensure his own safety while POK purged his board. Betty concludes her session by exporting these findings to her supervisor.

\section{Discussion}
\label{sec:discussion}

In this section, we discuss gaps in studying KG sensemaking as well as the trade-offs in using LLM-driven ontolgies. Finally, we discuss limitations and future work.

\subsection{Toward Knowledge Graph Sensemaking}
\label{sec:discussion/kg_sensemaking}
Our work foregrounds an open problem: it remains unclear how sensemaking actually proceeds over a knowledge graph.
Classic models such as Pirolli and Card describe sensemaking over documents as a foraging-and-synthesis loop \cite{Pirolli:2005:SensemakingProcess}, but we do not know whether knowledge graph analysis follows the same loop, a variant of it, or a different process altogether.
In designing Semantic Bundling, we found ourselves working both \textbf{top-down} (starting from a question and seeking supporting structure) and \textbf{bottom-up} (browsing the graph for what stands out) and often moving between the two.
Therefore, we offer Semantic Bundling and AgentK as instruments for studying this question.
Characterizing the knowledge graph sensemaking process, including its stages and how it differs from text document sensemaking, is important future work.

\subsection{Trade-offs of LLM-Defined Ontologies}
\label{sec:discussion/ontology}
To construct a KG from raw documents, AgentK lets an LLM define the ontology rather than building the graph against a fixed schema (Sect.~\ref{sec:agentk/visualizing}), expressing entities and relationships in open vocabulary and then operating over that ontology.
This was not a premise of our technique but a design choice that emerged in building AgentK, and it carries trade-offs worth examining.
Our usage scenarios (Sect.~\ref{sec:usage_scenarios}) suggest where this helps.
Open-vocabulary predicates yield human-readable labels and summaries that let a user grasp what a region of the graph is about without reading the underlying documents, and the same pipeline applies across datasets without schema engineering.
On the other hand, an LLM-defined ontology is less stable than a fixed schema.
The same relationship may be phrased many ways, and the graph inherits the quality of extraction and coreference resolution.
We layer a small fixed vocabulary of entity and relationship \emph{types} for encoding, which reintroduces a measure of rigidity, and our semantic deduplication can merge or separate relationships incorrectly, with errors propagating into the bundles built on top.
Between the two ends of the spectrum, an LLM-defined ontology trades formal precision for semantic richness; where the goal is human understanding rather than exact query, we found the trade worthwhile, but it warrants further study.

\subsection{Limitations and Future Work}
\label{sec:discussion/limitations}
We validate Semantic Bundling through use cases rather than a controlled study.
Whether the technique measurably improves cognitive processes, and how it fares over longer investigations, remains underexplored.
Our scope is closed document sets on the order of 1000 documents; scaling to substantially larger or continuously updated corpora raises open questions of graph layout, path-enumeration cost, and LLM latency that we have not yet characterized quantitatively.
Semantic Bundling also relies on an LLM to construct the KG when only source text is available.
While this import step is a convenience rather than a contribution (i.e., an existing pipeline or graph database can supply the KG instead), the quality of the extracted graph bounds everything built on top of it, and we leave a systematic comparison against dedicated KG-construction tools (e.g. Neo4j \cite{Neo4jKGBuilder}) to future work.

Our designs also inspire future explorations of visual analytics tools for knowledge graph sensemaking.
Many knowledge graphs have a temporal component and evolve over time, which would require extending bundles to compare graphs across time.
Further, users often work across multiple knowledge graphs, suggesting future operations that could link and reconcile entities and relationships between entirely separate KGs.
AgentK currently supports single-step undo but not a full history of an analysis.
Richer provenance tools could help users review and steer the bundling process.
Finally, the user invokes and parameterizes Semantic Bundling operations.
Composing these operations into agentic workflows that a user directs and verifies is a natural next step, provided it preserves the traceability that grounds the technique (Sect.~\ref{sec:sb/grounding}).

\section{Conclusion}
\label{sec:conclusion}

Making sense of large document collections as knowledge graphs can unlock new analytic capabilities by exposing relationships as an interpretable graph structure.
Yet this process often leaves users facing visual ``hairballs'' with unclear semantic meaning.
In this work, we presented \textbf{Semantic Bundling}, a visual analytics technique that uses LLMs to transform a KG into higher-level visual structures a user can read directly.
Through two operations---\textit{super nodes}, which collapse and summarize a region of the graph, and \textit{super edges}, which summarize the connection between two entities---Semantic Bundling makes relationships legible while keeping every summary traceable to the triples and source documents beneath it.
We realized the technique in \textbf{AgentK}, an open-source proof-of-concept system, and demonstrated through use cases how users uncover and verify insights with the source documents a traceable last resort rather than the starting point.
As people increasingly turn documents into knowledge graphs, techniques that make those graphs legible while keeping them grounded in evidence will be essential.
We see Semantic Bundling as a step toward KG sensemaking; an emerging practice we hope this work helps shape.

\bibliographystyle{IEEEtran}
\bibliography{main}

\end{document}